\documentclass[11pt,twoside,a4paper]{article}

\usepackage{graphicx}

\newcommand{\Tr}{\makebox{ Tr }}

\newcommand{\GeV}{\makebox{ GeV}}

\newcommand{\fm}{\makebox{ fm}}

\newcommand{\beq}{\begin{equation}}

\newcommand{\enq}{\end{equation}}

\newcommand{\beqa}{\begin{eqnarray}}

\newcommand{\enqa}{\end{eqnarray}}

\newcommand{\nn}{\nonumber}

\newcommand{\labelm}[1]{\label{#1}}

\newcommand{\lbfi}[1]{\labelm{#1}\end{figure}}

\newcommand{\lbq}[1]{\labelm{#1}\enq}

\newcommand{\lbqa}[1]{\labelm{#1}\enqa}

\newcommand{\befi}[2]{\begin{figure}[ht] \leavevmode \begin{center} \includegraphics[width=#2]{#1.eps} \end{center}}

\newcommand{\betwofi}[5]{
\begin{figure}[ht]
\leavevmode
\begin{minipage}{#5}
\includegraphics[width=#5]{#1.eps}
\end{minipage}
\hfill
\begin{minipage}{#5}
\includegraphics[width=#5]{#3.eps}
\end{minipage}\\
\begin{minipage}{#5}
\begin{center}
\small #2
\end{center}
\end{minipage}
\hfill
\begin{minipage}{#5}
\begin{center}
\small #4
\end{center}
\end{minipage}
}

\newcommand{\eq}[1]{eq.(\ref{#1})}

\newcommand{\fig}[1]{fig.(\ref{#1})}

\newcommand{\lbcap}[3]{\begin{center}\begin{minipage}{#1}\caption{\small #2}\labelm{#3}\end{minipage}\end{center}\end{figure}}

\newcommand{\lbtab}[3]{\centering\begin{minipage}{#1}\caption{\small #2}\labelm{#3}\end{minipage}\end{table}}

\newcommand{\pa}{\partial}

\newcommand{\cO}{\mbox{$\cal O$}}

\newcommand{\cP}{\mbox{$\cal P$}}

\newcommand{\cS}{\mbox{$\cal S$}}

\newcommand{\cN}{\mbox{$\cal N$}}

\newcommand{\bA}{\mbox{\bf A}}

\newcommand{\bF}{\mbox{\bf F}}

\newcommand{\bV}{\mbox{\bf V}}

\newcommand{\bW}{\mbox{\bf W}}

\newcommand{\bo}{\mbox{\bf 1}}

\newcommand{\al}{\alpha}

\newcommand{\ga}{\gamma}

\newcommand{\de}{\delta}

\newcommand{\ep}{\epsilon}

\renewcommand{\th}{\theta}

\newcommand{\ka}{\kappa}

\newcommand{\rh}{\rho}

\newcommand{\si}{\sigma}

\newcommand{\ch}{\chi}

\newcommand{\De}{\Delta}

\newcommand{\Ps}{\Psi}

\begin{document}
\title{\begin{flushright}\normalsize
TAUP-2502-98\\
HD-THEP-98-22\end{flushright}
\LARGE \bf Odd C-P contributions to diffractive processes}
\author{
{\Large Michael Rueter}\thanks{supported by a MINERVA-fellowship}\\[.3cm]
\itshape School of Physics and Astronomy\\
\itshape Department of High Energy Physics\\
\itshape Tel-Aviv University\\
\itshape 69978 Tel-Aviv, Israel\\
\itshape e-mail: {\tt rueter@post.tau.ac.il}\\[.5cm]
{\Large H.G.~Dosch, O.~Nachtmann}\\[.3cm]
\itshape Institut f\"ur Theoretische Physik\\
\itshape Universit\"at Heidelberg\\
\itshape Philosophenweg 16, D-69120 Heidelberg, FRG\\
\itshape e-mail: {\tt H.G.Dosch@ThPhys.Uni-Heidelberg.DE}\\
\itshape e-mail: {\tt O.Nachtmann@ThPhys.Uni-Heidelberg.DE}
}
\date{}
\maketitle
\thispagestyle{empty}
\begin{abstract}
We investigate contributions to diffractive scattering, which are odd under $C$- and $P$-parity. Comparison of $p$-$\bar p$ and $p$-$p$ scattering indicates that these odderon contributions are very small and we show how a diquark clustering in the proton can explain this effect. A good probe for the odderon exchange is the photo- and electroproduction of pseudo-scalar mesons. We concentrate on the $\pi^0$ and show that the quasi elastic $\pi^0$-production is again strongly suppressed for a diquark structure of the proton whereas the cross sections for diffractive proton dissociation are larger by orders of magnitude and rather independent of the proton structure.
\end{abstract}
\newpage
\setcounter{page}{1}
\section{Introduction}
In this paper we discuss those exchange contributions to diffractive
scattering which are odd under parity and charge conjugation. These
odd $C$ and $P$ contributions can have a similar energy dependence as the $C$ and $P$
even parts (the pomeron) and have been called odderon contributions
\cite{Lukaszuk:1973}. In approaches to high-energy scattering based on
QCD a sizable odd $C$ and $P$ contribution occurs naturally since
perturbative or nonperturbative three gluon exchange contains a part
which is odd under $C$ and $P$
\cite{Donnachie:1991}-\nocite{Nachtmann:1991}\nocite{Gauron:1993}\nocite{Armesto:1997}\nocite{Struminskii:1994}\nocite{Korchemsky:1996}\nocite{Wosiek:1997}\nocite{Gauron:1995}\nocite{Contogouris:1994}\cite{Rueter:1996}. A
comparison of the proton-antiproton data of the UA4/2-collaboration
\cite{Augier:1993} and an analysis of proton-proton data indicates that
the odderon contribution to proton-(anti)proton scattering is very
small (see e.g.~\cite{Bourrely:1984}). It has been pointed out by us
\cite{Rueter:1996,Rueter:1996III} that clustering of quarks inside the
nucleon is a plausible explanation for that suppression.

Recent experiments at HERA \cite{Tapprogge:1996}, following suggestions made in reference \cite{Schafer:1991}, are suited to extract
specifically the odderon contribution from high-energy photoproduction
amplitudes if suitably analyzed \cite{Kilian:1997}. In this paper we
discuss odderon contributions to general diffractive processes. Though
the underlying model is a special one, that of the stochastic vacuum
\cite{Dosch:1987,Dosch:1988}, here we concentrate on aspects which are
independent of the details of the model. Our paper is organized as
follows: In section 2 we give the general procedure of our approach
and some model independent results for the odderon contribution. In
section 3 we apply our results to photo- and electroproduction of
pseudoscalar mesons and point out experimental consequences. After the
summary we present in an appendix a technical derivation of a kind of
non-Abelian multipole expansion which is essential for our
calculation.

\section{General structure of Wegner-Wilson loop scattering}

We first shortly sketch some essential steps of our nonperturbative
treatment of high-energy scattering, for a detailed discussion we
refer to the original literature and reviews \cite{Dosch:1994}-\nocite{Dosch:1997III}\cite{Nachtmann:1996}. In a
general analysis of soft high-energy scattering Nachtmann
\cite{Nachtmann:1991,Nachtmann:1996} has evaluated the high-energy
quark-quark scattering amplitude in the femto-universe using the
eikonal approximation for the interaction of the quarks with the gluon
field. In a first step, we follow the same approach, and consider the
scattering amplitude of a very high energetic single quark in an
external gluon field $\bA_\mu$. Along its path $\Gamma$, the quark
picks up the eikonal phase $\bV$ (which here is a unitary $N_{\rm
 C}\times N_{\rm C}$ matrix)
\beq \bV=P\exp[-ig \int_\Gamma\bA_\mu(z)\ dz^\mu].
\lbq{3A1}
Here $\bA_\mu$ is the Lie-algebra valued potential and $P$ denotes
path ordering. The phase factor for an antiquark is obtained by
complex conjugation.

From the scattering amplitudes for single quarks in the background
field, 
we obtain the nonperturbative quark-quark scattering amplitude by
functional integration, with respect to the background field, of the 
product of the two quark scattering amplitudes. More specifically,
consider two quarks traveling along the lightlike paths $\Gamma_1$ and
$\Gamma_2$ given by 
\beq
\Gamma_1=(x^0,\vec b/2, x^3=x^0)~{\rm{and}}~\ \Gamma_2= (x^0, -\vec b/2, x^3=
-x^0),
\lbq{3A2}
corresponding to quarks moving  with velocity of light in opposite
directions, with an impact parameter $\vec b$ in the $x^1$-$x^2$-plane
(referred to in the following as the transverse plane). Let 
$\bV_{1,2}(\pm\vec b/2)$ be the phases picked up by the quarks along these paths
\beq
\bV_{1,2}(\pm\vec b/2)=P\exp\left[-ig \int_{\Gamma_{1,2}} \bA_\mu
(z)\ dz^\mu\right]. 
\lbq{3A3}
Then according to Nachtmann \cite{Nachtmann:1991} the scattering amplitude for two quarks with momenta $p_1$, $p_2$
and colors $c_1$, $c_2$ leading to two quarks with momenta $p_3$, $p_4$ and
colors $c_3\ c_4$ is given by 
\beq
T_{c_3c_4;c_1c_2}(s,t)=i\bar u (p_3)\gamma^\mu u(p_1)\bar u(p_4)\gamma_\mu u(p_2)~{\cal V}, 
\lbq{3A4}
where 
\beq
{\cal V}=-\big<Z^{-2}_\psi\big>_A \big< \int d^2 b\ e^{-i\vec \De_\perp \cdot
\vec b}\{ [ \bV_1(-\vec b/2)-\bo ]_{c_3c_1}
[\bV_2 (+\vec b/2)-\bo ]_{c_4c_2}\}\big> _A. 
\lbq{3A5}
Here $< >_A$ denotes functional integration over the
background field; $\vec \De_\perp$ is the momentum transfer $(p_1-p_3)$ projected
on the transverse plane. Of course the approximation makes sense only  
if $|\vec \De_\perp|$ is much smaller then the momentum of the quarks. The quantity $Z_\psi$ is the fermion wavefunction 
renormalization constant in the eikonal approximation, given by
\beq
Z_\psi[A]={1\over N_{\rm C}}\Tr \{\bV_1(0)\}={1\over N_{\rm C}} {\Tr} \{\bV_2(0)\}.
\lbq{3A6}
The subtraction of the unit operator from the phase-matrices $\bV$ is due
to the transition from the $S$- to the $T$-operator.

In the limit of high energies we have  helicity conservation
\beq
\bar u(p_3)\gamma^\mu u(p_1)\bar u(p_4)\gamma_\mu u(p_2)
\mathop{\longrightarrow}_{s \to \infty} 2s\delta_{\lambda_1\lambda_3}
\delta_{\lambda_4 \lambda_2}
\lbq{3A7}
where $\lambda_i$ are the helicities of the quarks and $s=(p_1+p_2)^2$.
In the following we can thus ignore the spin degrees of freedom.

The scattering amplitude (\eq{3A4}) is explicitly gauge dependent.
But in hadron-hadron scattering the constituents form color-neutral
objects: an antisymmetric three quark state for the baryons and a
quark-antiquark state for mesons. The constituents move in the
femto-universe on nearly parallel lightlike lines. We consider first
the simpler case of the meson which is modeled as a superposition of
colorless dipoles the size distribution of which is given by the
transversal wavefunction. The local gauge invariant colorless
quark-antiquark dipoles are represented in space-time as Wegner-Wilson
loops whose lightlike sides are formed by the quark and antiquark
paths, and front ends by the Schwinger strings ensuring local gauge
invariance (see \fig{vac2}).
\befi{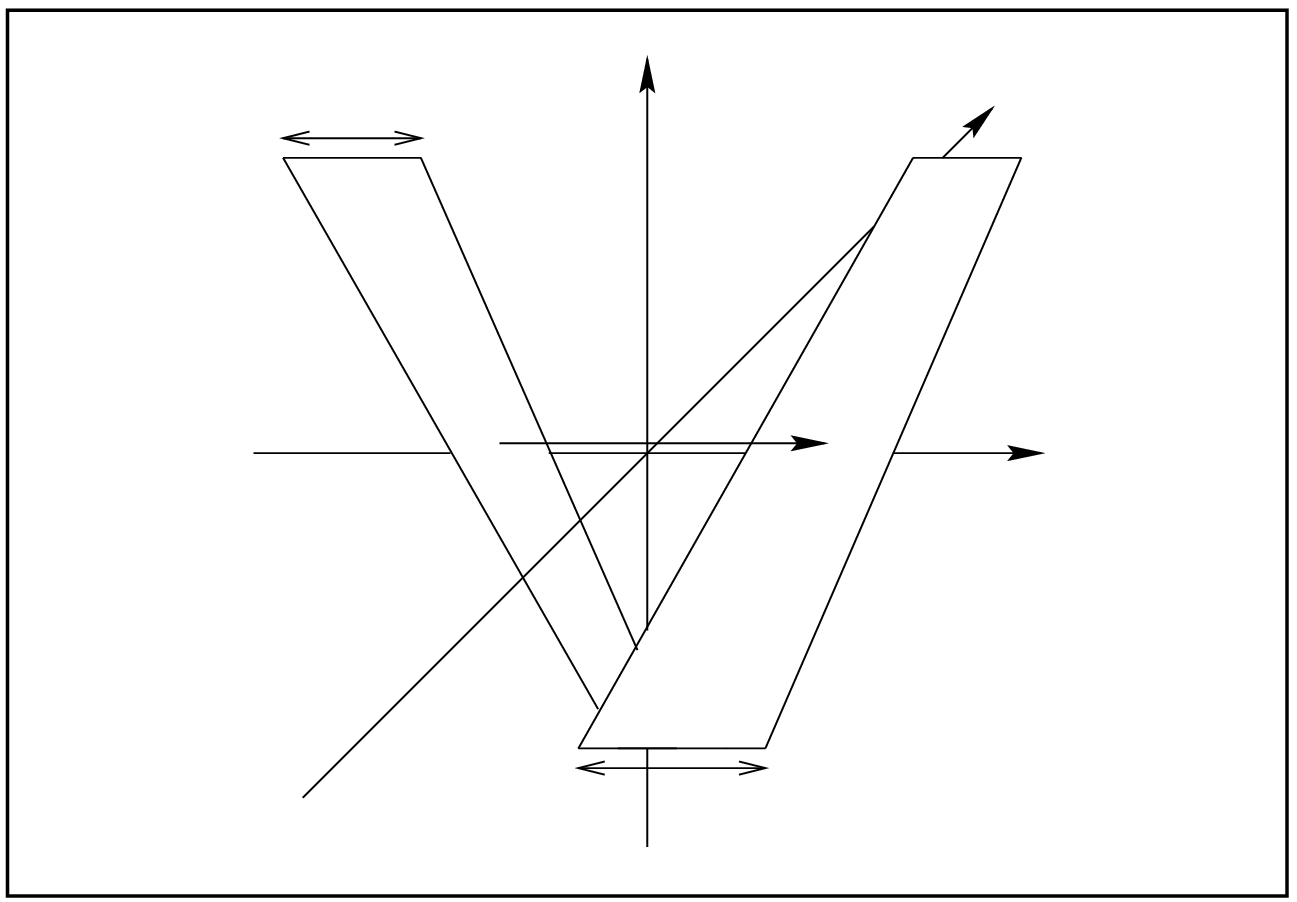}{8.5cm}
\unitlength.85cm
\begin{picture}(0,0)
\small
\put(5.4,5.4){loop 2}
\put(12.4,5.4){loop 1}
\put(12.2,3.9){$\vec{x}$}
\put(9.5,7.1){$x^0$}
\put(12.2,6.9){$x^3$}
\put(6.9,7.0){$\vec{R}_2$}
\put(9.5,1.4){$\vec{R}_1$}
\put(9.0,4.6){$\vec{b}$}
\end{picture}
\lbcap{14cm}{Wegner-Wilson loops formed by the paths of quarks
and antiquarks inside two dipoles. The impact parameter $\vec b$ is the
distance vector between the middle lines of the two loops. $\vec R_1$ and
$\vec R_2$ are the vectors in the transverse plane from the quark
lines to the antiquark lines of dipole 1 and 2 respectively. The front lines
of the loops guarantee that the dipoles behave as singlets under local
gauge transformations.}{vac2}

The resulting loop-loop amplitude is now specified not only by the
impact parameter, but also by the transverse extension vectors of the
loops. We thus introduce the dipole-dipole rather than a quark-quark
profile function:
\beq
\tilde J (\vec b,\vec R_1,\vec R_2)=
\frac{-\big< W_1 W_2\big> _A}{ \big< {1\over N_{\rm C}} \Tr \bW_1(0,\vec R_1)\big> _A
\big< {1\over N_{\rm C}}\Tr\bW_2(0,\vec R_2)\big>_A},
\lbq{3A8}
where the path $\partial S_1$ of the closed Wegner-Wilson loop $\bW[\cS_{1}]$ in 
\beq
W_i=\frac{1}{N_{\rm C}}\Tr \left\{\bW[\cS_{i}]-\bo\right\}
\lbq{WWloopb}
is a rectangle whose long sides are formed by the quark path
$\Gamma_1^q=(x^0,\vec b/2+\vec R_1/2,x^3=x^0)$ and the antiquark path
$\Gamma_1^q=(x^0,\vec b/2-\vec R_1/2,x^3=x^0)$ and whose front sides are
formed by lines from $(T,\vec b/2+\vec R_1/2,T)$ to $(T,\vec b/2-\vec
R_1/2,T)$ for large positive and negative $T$ (we will then take the
limit $T\to \infty$). $W_2$ is constructed
analogously. The denominator in \eq{3A8} is the loop
renormalization that replaces the quark field renormalization in \eq{3A5}.
Meson-meson scattering is obtained from dipole-dipole
scattering by smearing over the transversal wavefunctions:
\[
J_{\rm MM}(\vec b) = \int \frac{d^2R_1}{4\pi}\frac{ d^2 R_2}{4\pi} \tilde J(\vec b,\vec R_1,\vec R_2) \Ps_i(\vec R_1)
\Ps_{i'}^*(\vec R_1)\Ps_{k}(\vec R_2)\Ps_{k'}^*(\vec R_2)
\]
where $i,k,i'$ and $k'$ represent the in- respectively out-coming mesons. For the elastic case we have a density and for the inelastic case an overlap function.

For baryons the constituting ``tripole'' of quarks is made locally gauge invariant by three Schwinger strings starting from the quarks and being connected anti-symmetrically at some point $y$. The world lines of the three quarks and the $y$-point together with  the Schwinger strings form thus the product of three Wegner-Wilson loops (without traces) with one common line, that of the $y$-point: 
\beqa
B_i&=&\frac{1}{6}\ep_{abc}\ep_{a'b'c'}\left\{ \bW_{a'a}[\cS_{i1}]\bW_{b'b}[\cS_{i2}]\bW_{c'c}[\cS_{i3}]-\de_{a'a}\de_{b'b}\de_{c'c}\right\}.
\lbqa{Jtilde}
The unitary $3\times 3$ matrices $\bW [\cS_{ij}]$ are Wegner-Wilson loops (without traces)
\beq
\bW_{a'a}[\cS_{ij}]= \left[\cP\, e^{-ig\oint_{\pa S_{ij}}\bA_\mu (z)\,dz^\mu}\right]_{a'a}
\lbq{WWloopa}
and the integration paths $\pa \cS_{ij}$ are illustrated in \fig{3loops}. The expression $B_i$ has to be inserted for $W_i$ in \eq{3A8} if a baryon is considered instead of a meson.
\befi{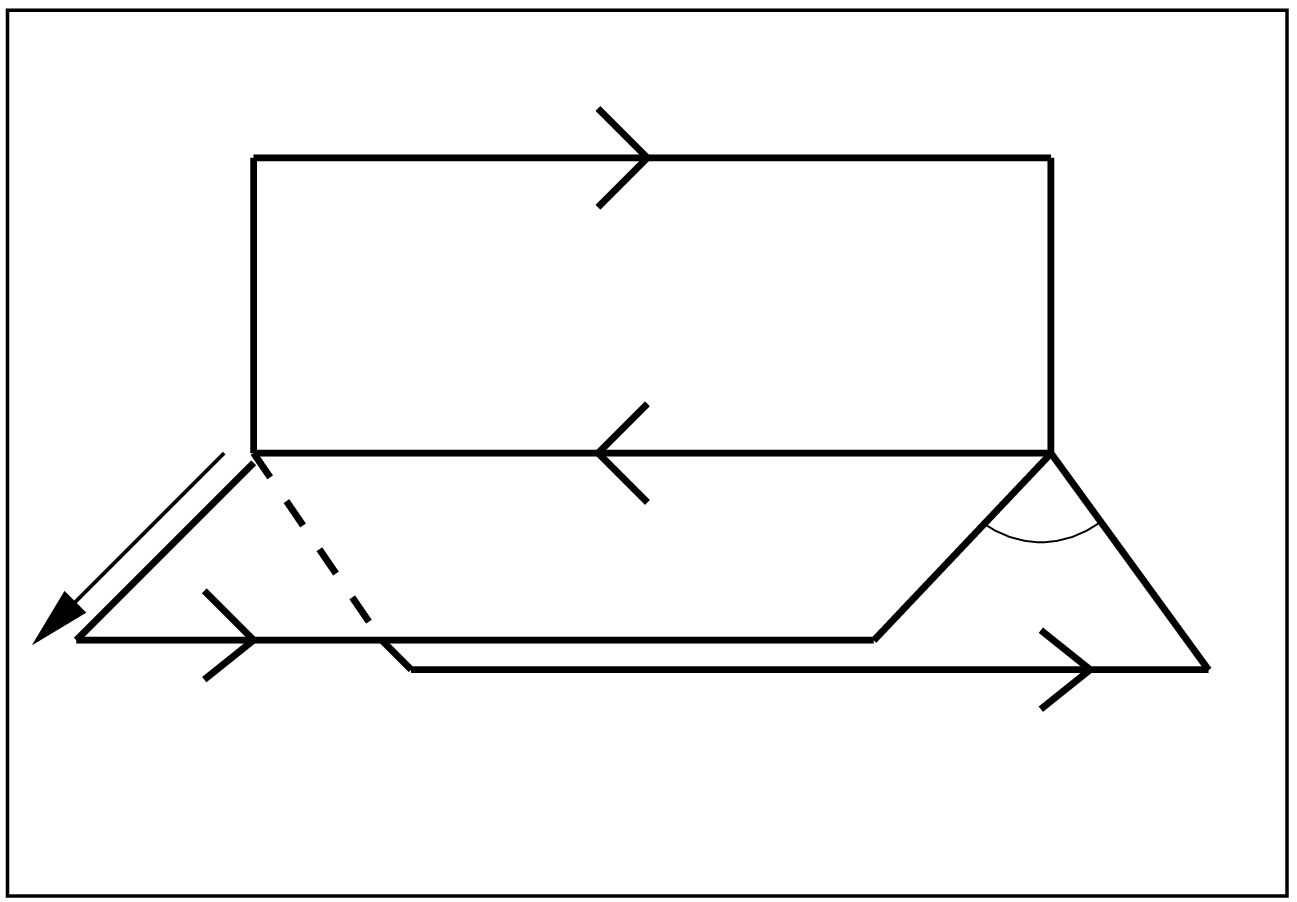}{6cm}
\unitlength1cm
\begin{picture}(0,0)
\small
\put(9.7,2.6){$\al$}
\put(9.7,1.4){$\pa \cS_{i2}$}
\put(5.7,1.6){$\pa \cS_{i3}$}
\put(6.7,4.5){$\pa \cS_{i1}$}
\put(5.0,2.8){$\vec{R}_i/2$}
\end{picture}
\begin{center}\begin{minipage}{14cm}\caption{\small The colorless $qqq$-object is constructed out of three Wegner-Wilson loops with one common line which transforms like a color-singlet. Here $\pa \cS_{ij}$ denotes the loop corresponding to quark j of \mbox{baryon i}. By varying the angle $\al$ we can consider different geometries of the baryon. With $\vec{R}_i$ we denote the extension of the baryon. For the calculation the length of the lightlike paths has to be taken as infinity.}\labelm{3loops}\end{minipage}\end{center}\end{figure}

If the angle $\al$, introduced in \fig{3loops}, tends to zero we have a baryon constructed out of a quark and a diquark. In reference \cite{Rueter:1996} we showed that for vanishing diquark size this configuration is exactly equivalent to a meson built from one Wegner-Wilson loop:
\[
B_i\mathop{\Rightarrow}_{\al\to 0} W_i.
\]
The profile function (\eq{3A8}) can be written generically as:
\beq
\tilde{J}=-\big< O_1\times O_2\big> ,
\lbq{redampl}
where $O_i$ is either $B_i$ or $W_i$ and the brackets denote functional integration over the gluon fields. To obtain the scattering amplitude at center of mass energy $s$ and momentum transfer \mbox{$t=-\vec{\De}_\perp^2$} for fixed Wegner-Wilson loop constructions one has to integrate over the impact parameter $\vec{b}$
\beq
T(s,t) = 2is \int \, d^2b\,e^{-i\vec{\De}_\perp\cdot\vec{b}}\,\tilde{J}.
\lbq{T}
For the total cross section follows:
\beq
\si^{\rm tot}=\frac{1}{s}{\rm Im}T(s,0)
\lbq{sigtotundslope}
which is independent of the center of mass energy $s$. In order to account for the slight energy dependence observed for hadronic and photoproduction cross sections one can introduce an energy dependent hadron radius (see reference \cite{Dosch:1994}) which can reproduce the $p$-$\bar p$ data very well. Unless otherwise stated we have adjusted the proton radius in this paper to a cm-energy of $W= 20\GeV$. Though the energy dependence of hadronic cross sections is a very interesting effect, it plays only a minor role in this investigation where we look for effects which change the cross sections by orders of magnitude.

When we consider physical reactions the profile function (\eq{redampl}) is smeared with wavefunctions for the size and orientation of the loops which can be either of phenomenological nature (mesons, baryons) or perturbatively motivated (for example the dipole of the virtual photon in vectormeson electroproduction).

In a series of papers \cite{Rueter:1996,Rueter:1996III,Dosch:1994,Dosch:1997,Dosch:1997II,Rueter:1997II} the model of the stochastic vacuum (MSV) \cite{Dosch:1987,Dosch:1988,Simonov:1988} has been used to calculate diffractive hadron-hadron scattering, photo- and electroproduction of vectormesons and the $\De\rh$-parameter of proton-(anti)proton scattering. Using the MSV we approximate the integration over the slowly varying gluon field in \eq{redampl} with a Gaussian stochastic process using an ansatz for the nonlocal non-Abelian gluon condensate. The analytic form of this nonlocal condensate has been fitted to lattice results of this quantity \cite{DiGiacomo:1992,DiGiacomo:1996} and has mainly two parameters: the absolute value for vanishing separation, the usual gluon condensate $<g^2FF>$, and the characteristic length scale on which the nonlocal condensate falls off, the correlation length $a$.

In order to perform the integration over the gluon field we first transform the path integrals of the Wegner-Wilson loops into surface integrals over the parallel transported field strengths using the non-Abelian Stokes-theorem \cite{Bralic:1980} and
expand the exponentials, leading for dipoles to expressions like
\[
\big< \Tr (\bF \bF) \Tr(\bF \bF )\big> , \; \big< \Tr (\bF \bF \bF )\Tr(\bF \bF \bF )\big> ,\dots
\]
In the Gaussian approximation adopted in our model the product factorizes
into a product of nonlocal gluon condensates.

The integration over the lightlike coordinates of the surfaces can be
performed analytically and one ends up with integrations of the
transversal coordinates. A geometrical picture of the integration in
the transversal plane is given in \fig{trans}. The thick
lines from the central points to the (anti-)quarks are the transversal
projections of the surfaces created in applying the Stokes-theorem.
The term $\tilde\chi_{ij}$ represents the contribution of a correlator of a
field strength on the piece $i$ of hadron 1 with a field strength $j$
of hadron 2. For a baryon we have three projection lines, one from each
quark to the central point.
\befi{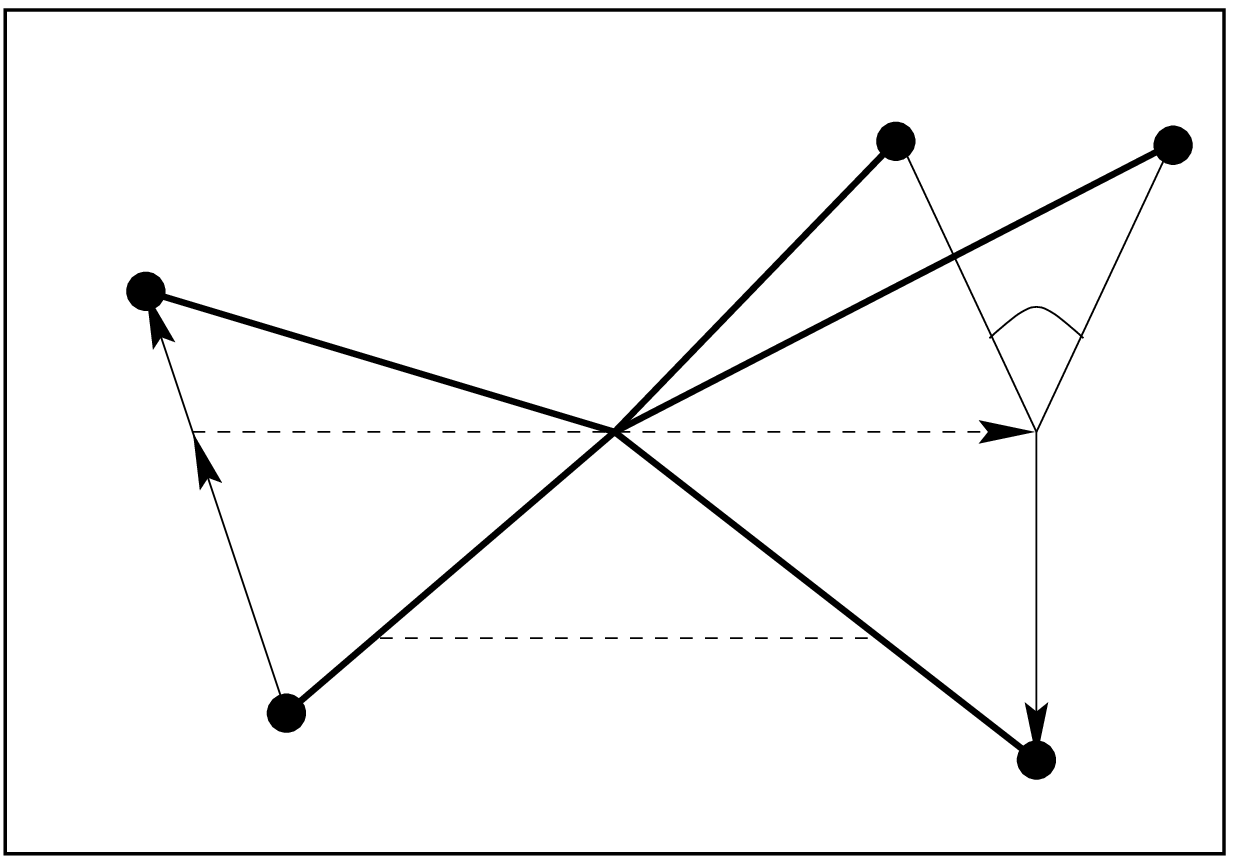}{8cm}
\unitlength1cm
\begin{picture}(0,0)
\small
\put(5.7,1.3){1}
\put(4.8,4.7){2}
\put(10.6,1.0){1}
\put(11.5,5.8){2}
\put(9.7,5.8){3}
\put(6.5,3.1){$\vec b$}
\put(4.2,3.7){$z\vec R_1$}
\put(4.7,2.5){$\bar z \vec R_1$}
\put(10.9,2.5){$\vec R_2/2$}
\put(10.6,4.0){$\alpha$}
\put(7.7,1.8){$\tilde\ch _{11}$}
\end{picture}
\lbcap{14cm}{A geometrical picture of the integration in the transversal plane. The constituents are denoted by the black dots. The two objects scatter off with impact parameter $\vec{b}$. A baryon is always constructed from three quarks with the same distance from the central point. We only vary the angle between two of them. The thick lines from the central point to the (anti-)quarks are the
transversal projections of the surfaces created in applying the
Stokes-theorem. The term $\tilde\chi_{ij}$ represents the contribution of
a correlator of a field strength on the piece $i$ of hadron 1 with a
field strength $j$ of hadron 2. The integration has to be performed
over all the transversal projections of the surface, i.e.~1 and 2 of
a meson combined with 1, 2 and 3 of a baryon. The impact parameter $\vec
b$ points to the lightcone barycenter of the dipole, i.e.~the
distance between the quark and antiquark is divided according to the
longitudinal momentum fraction of each constituent which is given by $z$ and $\bar{z}=1-z$ \protect\cite{Dosch:1997}.}{trans}

The real functions $\tilde{\ch}_{ij}$ depend only on the transversal coordinates and is given by \cite{Dosch:1994,Rueter:1997II}:
\beqa
\tilde{\ch}_{ij}&=&<g^2FF>\left( \ka \, \int_0^1dw_1\, \int_0^1dw_2\, \vec{r}_{1i}\cdot\vec{r}_{2j}\, f_1\left( |w_1 \vec{r}_{1i}-w_2\vec{r}_{2j}|\right)\right.\nn\\
&&\hspace{2cm}+ (1-\ka )f_2\left( |\vec{r}_{1i}-\vec{r}_{2j}|\right)\bigg).
\lbqa{chi}
The vector $\vec{r}_{1i}$ ($\vec{r}_{2j}$) points to constituent $i$ ($j$) of object 1 (2) and is a function of $\vec{b}$, $\vec{R}$ and $z$ as indicated in \fig{trans}. The usual gluon condensate is denoted by $<g^2FF>$ and the parameter $\ka$ and the two functions $f_1$ and $f_2$ depend on the explicit ansatz for the nonlocal gluon condensate and fall off on the length scale given by the correlation length $a$. In reference \cite{Dosch:1994} it was also shown that one of the $w$-integrations in \eq{chi} can be done analytically.

In the following we give the explicit result for the first two orders of the profile function in an expansion of the $O_i$ of \eq{redampl} in the field strengths. The results can be expressed in terms of the functions $\tilde{\ch}_{ij}$.

The leading non vanishing contribution to the profile function was calculated in reference \cite{Dosch:1994}. For dipole-dipole scattering it results from the second order expansion of $W_i$ in the field strengths. This contribution to the scattering amplitude is purely imaginary and even under charge parity. That's why we call this leading contribution the pomeron contribution:
\beq
\tilde{J}_{\rm DD}^{C=+1}=\frac{1}{8 N_{\rm C}^2(N_{\rm C}^2-1)12^2}\left( \tilde{\ch}_{11}-\tilde{\ch}_{12}-\tilde{\ch}_{21}+\tilde{\ch}_{22}\right)^2.
\lbq{ddcplus}
Looking at baryons instead of dipoles we have to expand $B_i$ instead of $W_i$ with the result:
\beqa
\tilde{J}_{\rm BB}^{C=+1}&=&\frac{1}{8 N_{\rm C}^2(N_{\rm C}^2-1)12^2}\left( \tilde{\ch}_{11}^2+\tilde{\ch}_{12}^2+\tilde{\ch}_{13}^2+\tilde{\ch}_{21}^2+\tilde{\ch}_{22}^2+\tilde{\ch}_{23}^2+\tilde{\ch}_{31}^2+\tilde{\ch}_{32}^2+\tilde{\ch}_{33}^2\right.\nn\\
&&-\tilde{\ch}_{11}\tilde{\ch}_{12}-\tilde{\ch}_{11}\tilde{\ch}_{13}-\tilde{\ch}_{12}\tilde{\ch}_{13}-\tilde{\ch}_{21}\tilde{\ch}_{22}-\tilde{\ch}_{21}\tilde{\ch}_{23}-\tilde{\ch}_{22}\tilde{\ch}_{23}-\tilde{\ch}_{31}\tilde{\ch}_{32}-\tilde{\ch}_{31}\tilde{\ch}_{33}\nn\\
&&-\tilde{\ch}_{32}\tilde{\ch}_{33}-\tilde{\ch}_{11}\tilde{\ch}_{21}-\tilde{\ch}_{11}\tilde{\ch}_{31}-\tilde{\ch}_{21}\tilde{\ch}_{31}-\tilde{\ch}_{12}\tilde{\ch}_{22}-\tilde{\ch}_{12}\tilde{\ch}_{32}-\tilde{\ch}_{22}\tilde{\ch}_{32}-\tilde{\ch}_{13}\tilde{\ch}_{23}\nn\\
&&-\tilde{\ch}_{13}\tilde{\ch}_{33}-\tilde{\ch}_{23}\tilde{\ch}_{33}\nn\\
&&+\frac{1}{2}\left( \tilde{\ch}_{11}\tilde{\ch}_{22}+\tilde{\ch}_{11}\tilde{\ch}_{23}+\tilde{\ch}_{11}\tilde{\ch}_{32}+\tilde{\ch}_{11}\tilde{\ch}_{33}+\tilde{\ch}_{12}\tilde{\ch}_{21}+\tilde{\ch}_{12}\tilde{\ch}_{23}+\tilde{\ch}_{12}\tilde{\ch}_{31}+\tilde{\ch}_{12}\tilde{\ch}_{33}\right.\nn\\
&&+\tilde{\ch}_{13}\tilde{\ch}_{21}+\tilde{\ch}_{13}\tilde{\ch}_{22}+\tilde{\ch}_{13}\tilde{\ch}_{31}+\tilde{\ch}_{13}\tilde{\ch}_{32}+\tilde{\ch}_{21}\tilde{\ch}_{32}+\tilde{\ch}_{21}\tilde{\ch}_{33}+\tilde{\ch}_{22}\tilde{\ch}_{31}+\tilde{\ch}_{22}\tilde{\ch}_{33}\nn\\
&&\left.+\tilde{\ch}_{23}\tilde{\ch}_{31}+\tilde{\ch}_{23}\tilde{\ch}_{32}\right)\Big).
\lbqa{bbcplus}
By putting for example quark 3 of baryon 1 on top of quark 2, that is $\tilde{\ch}_{3j}\Rightarrow\tilde{\ch}_{2j}$, we finally obtain the leading contribution to the profile function of dipole-baryon scattering:
\beqa
\tilde{J}_{\rm DB}^{C=+1}&=&\frac{1}{8 N_{\rm C}^2(N_{\rm C}^2-1)12^2}\left( \left(\tilde{\ch}_{11}-\tilde{\ch}_{21}\right)^2+\left(\tilde{\ch}_{12}-\tilde{\ch}_{22}\right)^2+\left(\tilde{\ch}_{13}-\tilde{\ch}_{23}\right)^2\right.\\
&&-\left(\tilde{\ch}_{11}-\tilde{\ch}_{21}\right)\left(\tilde{\ch}_{12}-\tilde{\ch}_{22}\right)-\left(\tilde{\ch}_{11}-\tilde{\ch}_{21}\right)\left(\tilde{\ch}_{13}-\tilde{\ch}_{23}\right)-\left(\tilde{\ch}_{12}-\tilde{\ch}_{22}\right)\left(\tilde{\ch}_{13}-\tilde{\ch}_{23}\right)\Big).\nn
\lbqa{dbcplus}
One important consequence of our result is that the profile function of a dipole is symmetric by rotating the dipole by $\pi$ and replacing $z$ by $\bar{z}$, that is $\tilde{\ch}_{i2}\Leftrightarrow\tilde{\ch}_{i1}$. This replacement just interchanges the quark with the antiquark and the $C=+1$ contribution has to be symmetric under this transformation. But this symmetry also applies for a baryon with a point like diquark.

Now we give the result for the next to leading order of the expansion in the field strength correlators of the profile function. It was calculated in reference \cite{Rueter:1996} but only the contribution which is odd under charge parity could be easily computed. This contribution gives the leading real part of the $C=-1$ scattering amplitude and for dipoles it is given by
\beq
\tilde{J}^{C=-1}_{\rm DD}=-i\frac{5}{8 N_{\rm C}^2(N_{\rm C}^2-1)^2 12^3 36} \left(\tilde{\chi}_{11}-\tilde{\chi}_{12}-\tilde{\chi}_{21}+\tilde{\chi}_{22}\right)^3.
\lbq{ddcminus}
For baryons we obtain analogously
\beqa
\tilde{J}^{C=-1}_{\rm BB}&=&-i\frac{5}{8N_{\rm C}^2 (N_{\rm C}^2-1)^212^3 36}\bigg( \tilde{\chi}_{11}^3\, + \, \mbox{8 permutations} - \frac{3}{2} \tilde{\chi}_{11}^2\,\tilde{\chi}_{12}\, + \, \mbox{35 per.}\nn\\
&&+\frac{3}{4} \tilde{\chi}_{11}^2\,\tilde{\chi}_{22}\, + \, \mbox{35 per.} \, + \,\frac{3}{2} \tilde{\chi}_{11}\,\tilde{\chi}_{12}\,\tilde{\chi}_{22}\, + \, \mbox{35 per.}+6 \tilde{\chi}_{11}\,\tilde{\chi}_{12}\,\tilde{\chi}_{13}\, + \, \mbox{5 per.}\nn\\
&&-3 \tilde{\chi}_{11}\,\tilde{\chi}_{12}\,\tilde{\chi}_{23}\, + \, \mbox{35 per.}  +6 \tilde{\chi}_{11}\,\tilde{\chi}_{22}\,\tilde{\chi}_{33}\, + \, \mbox{5 per.} \bigg).
\lbqa{bbcminus}
Here only one contribution to every different geometrical situation is given explicitly and the number of additional permutations is shown. For more details see reference \cite{Rueter:1996}. The dipole-baryon result is obtained in the same way as described above:
\beqa
\tilde{J}_{\rm DB}^{C=-1}&=&-i\frac{5}{8 N_{\rm C}^2(N_{\rm C}^2-1)^2 12^3 36}\left( \frac{1}{2}\left(\tilde{\ch}_{11}-\tilde{\ch}_{12}-\tilde{\ch}_{21}+\tilde{\ch}_{22}\right)^3+\frac{1}{2}\left(\tilde{\ch}_{11}-\tilde{\ch}_{13}-\tilde{\ch}_{21}+\tilde{\ch}_{23}\right)^3\right.\nn\\
&&+\frac{3}{2}\left(-2\tilde{\ch}_{11}+\tilde{\ch}_{12}+\tilde{\ch}_{13}+2\tilde{\ch}_{21}-\tilde{\ch}_{22}-\tilde{\ch}_{23}\right)\left(\tilde{\ch}_{12}-\tilde{\ch}_{13}-\tilde{\ch}_{22}+\tilde{\ch}_{23}\right)^2\bigg).
\lbqa{dbcminus}
These results are antisymmetric under the exchange of $\tilde{\ch}_{i2}\Leftrightarrow\tilde{\ch}_{i1}$ as it should be for the $C=-1$ contribution. But this is again also true for a baryon with a point like diquark and this has the following important consequences: If the proton  can be described as a quark-diquark system with a point like diquark then scattering amplitudes with odd $P$ exchange and without proton breakup vanish. The amplitude will in general not vanish if the proton is broken up, since the final state may contain states with parity opposite to that of the incoming proton and the overlap function contains in general antisymmetric contributions.  

The assumption of a strictly point like diquark is of course
completely unrealistic. In order to get an analytic expression for the
suppression with small diquark sizes we expand the profile function
(\eq{dbcminus}) in the diquark extension $d$ and keep only those terms
which survive if integrated over with a (symmetric) baryon density. The result is derived in the appendix:
\beqa
\tilde{J}_{\rm DB}^{C=-1}&=
&-i\frac{5}{8 N_{\rm C}^2(N_{\rm C}^2-1)^2 12^3 36}\nn\\
&&\times\left( (-3)\left(\tilde{\ch}_{11}-\tilde{\ch}_{12}-\tilde{\ch}_{21}+\tilde{\ch}_{22}\right)\left(\tilde{\ch}_{12}-\tilde{\ch}_{13}-\tilde{\ch}_{22}+\tilde{\ch}_{23}\right)^2\right)\nn\\
&&+\cO(d^3).
\enqa
The second term ($\left(\tilde{\ch}_{12}-\tilde{\ch}_{13}-\tilde{\ch}_{22}+\tilde{\ch}_{23}\right)^2$) is of order $d^2$ resulting in differential cross sections which vanish with the diquark size like $d^4$.
\section{Application to physical reactions}
\labelm{secreactions}
\subsection{Proton-proton and proton-antiproton scattering}
The quantity most sensitive to an odd $C$ and $P$ exchange is $\Delta \rho$, the difference between the ratio of real to imaginary part of the scattering amplitude for $p$-$p$ and $\bar p$-$p$ scattering. We here only quote the results from reference \cite{Rueter:1996,Rueter:1996III} and show in \fig{rhoplot} the diagram for the dependence of
\beq
\De \rh= \rh^{\bar p p} - \rh^{p p}=\frac{{\rm Re}\left[ T^{\bar p p}\right]}{{\rm Im}\left[  T^{\bar p p}\right]}-\frac{{\rm Re}\left[ T^{ p p}\right]}{{\rm Im}\left[ T^{ p p}\right]}
\lbq{rhodef}
on the diquark size $d$.
\befi{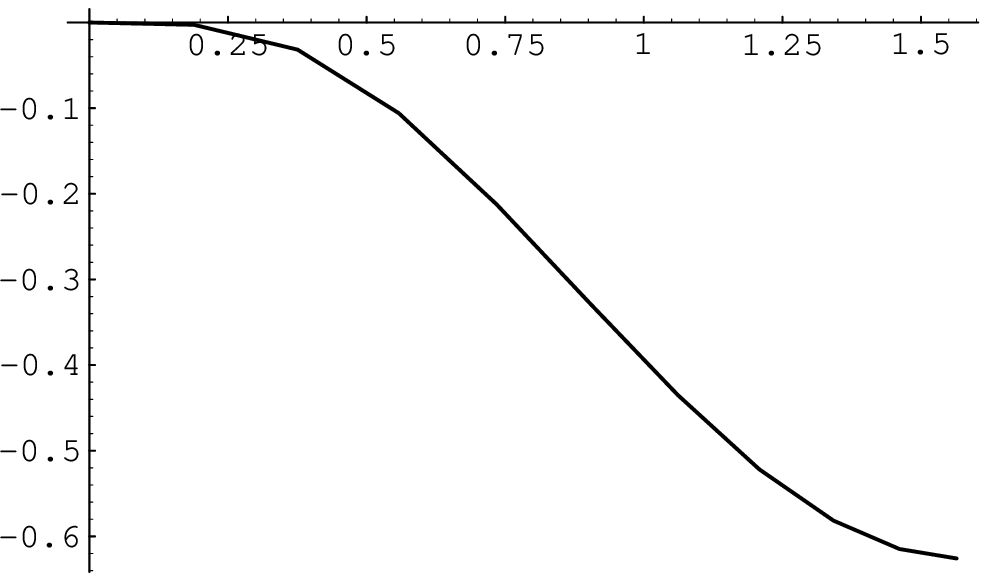}{7.5cm}
\unitlength1cm
\begin{picture}(0,0)
\put(5.1,1){$\De \rh$}
\put(10.5,4.5){$d$ [fm]}
\end{picture}
\lbcap{14cm}{$\De \rh$ at UA4/2 energy for proton-(anti)proton scattering as a function of the diquark size $d$.}{rhoplot}
\\As can be seen from \fig{rhoplot} a clustering of two quarks to a diquark with an extension smaller or equal to $0.3$ fm yields a drastic suppression of $\De \rh$ to a value $|\De \rh | \le 0.02$ which is compatible with the analysis of experiments.
\subsection{Elastic electroproduction of $\pi^0$}
Since the $\pi^0$ has $C$-parity +1 the photon-pion overlap function is 
antisymmetric and couples only to the odderon. Therefore in order
to calculate the amplitude for the reaction $\ga p \rightarrow \pi^0 p$ we have to smear the profile function $\tilde{J}_{\rm DB}^{C=-1}$ (\eq{dbcminus}) with appropriate wavefunctions. In our frame-work it is natural to use light cone wavefunctions \cite{Bjorken:1971}-\nocite{Gunion:1977}\cite{Lepage:1980}. The photon can be described by a quark-antiquark dipole of flavor $f$ with helicities $h_1$ and $h_2$ and light cone momentum $z$. We obtain for the profile function \cite{Brodsky:1994,Dosch:1997}:
\beq
J=\int \frac{d^2 r_{\rm P}}{4\pi}\int \frac{d^2 r_\ga}{4\pi}dz\sum_{f,h_1, h_2}\Psi^{*\, \pi^0}_{f h_1 h_2}(\vec{r}_\ga,z)\Psi^{\ga}_{f h_1 h_2}(\vec{r}_\ga,z)\left|\Psi^{\rm P}(r_{\rm P})\right|^2\times \tilde{J}.
\lbq{profilepi}
The scattering amplitude is then given by integrating over the impact parameter (\eq{T}) and for the differential elastic cross section we obtain:
\[
\frac{d\si^{\rm el}}{dt}=\frac{1}{16\pi s^2}|T|^2.
\]
For the proton wavefunction we use an simple Gaussian ansatz
\[
\Psi^{\rm P}(r_{\rm P})=\frac{\sqrt{2}}{S_{\rm P}}e^{-\frac{r_{\rm P}^2}{4S_{\rm P}^2}}.
\]
All the wavefunctions are normalized as follows:
\[
\int \frac{d^2 r}{4\pi}dz\left|\Psi(r,z)\right|^2=1,
\]
where for the photon and pion also a sum over flavor and helicity is included. The photon wavefunctions can be computed using light cone perturbation theory \cite{Bjorken:1971,Lepage:1980} and are given for our normalization in reference \cite{Dosch:1997}. They depend on the polarization and virtuality of the photon. In reference \cite{Dosch:1997II} the application was extended to real photons by using (anti)quark masses that depend on the virtuality and become equal to the constituent masses for $Q^2=0$, $m(Q^2=0)=220$ MeV. The pion wavefunction is parametrized similarly. The spin structure is that of the pseudo-scalar current of quarks with mass $m(Q^2)$. The distribution of the transverse and longitudinal momentum is discussed below (in the following we work in momentum space):
\beq
\Psi^{\pi^0}_{f h_1 h_2}(\vec{p},z)=c_f\Psi^{\pi^0}_{h_1 h_2}(\vec{p},z)=c_f\frac{1}{\sqrt{2}}\left(\de_{+-}-\de_{-+}-\frac{p^1-ip^2}{m}\de_{++}-\frac{p^1+ip^2}{m}\de_{--}\right)\Psi^{\pi^0}(p,z),
\lbq{piwf1}
where for example $\de_{+-}=\de_{h_1+}\de_{h_2-}$ and $c_f=\pm\frac{1}{\sqrt{2}}$ for $f=u,d$. Normalization of the pion state gives
\beq
\int \frac{d^2 p}{16\pi^3}dz\left( 1+\frac{\vec{p}^2}{m^2}\right)\left|\Psi^{\pi^0}(p,z)\right|^2=1
\lbq{piwf2}
and comparison with $\pi\rightarrow\mu\nu$ and $\pi\rightarrow \ga\ga$ (using PCAC) gives two more normalization conditions (see for example \cite{Lepage:1981}):
\beqa
\int \frac{d^2 p}{16\pi^3}dz\Psi^{\pi^0}(p,z)&=&\frac{f_\pi}{2\sqrt{N_{\rm C}}}\\
\int dz\Psi^{\pi^0}(p=0,z)&=&\frac{\sqrt{N_{\rm C}}}{f_\pi}.
\lbqa{piwf3}
For $\Psi^{\pi^0}(p,z)$ we make an ansatz which has an exponential dependence on $p$ and a $z$ dependence modeled in the way proposed by Wirbel, Stech and Bauer \cite{Wirbel:1985}:
\beqa
\Psi^{\pi^0}(p,z)&=&\frac{\sqrt{N_{\rm C}}}{f_\pi}z(1-z)f(z)e^{-\frac{p^2}{2w^2}},\nn\\
f(z)&=&\cN \sqrt{z(1-z)}e^{-\frac{M_\pi^2(z-1/2)^2}{2w^2}}.
\lbqa{piwf4}
Using the conditions \eq{piwf2}-\eq{piwf3} we fix the parameters:
\beqa
w&=&\frac{2\pi}{\sqrt{N_{\rm C}}}f_\pi,\nn\\
\frac{1}{\cN}&=&\int dz (z(1-z))^{3/2} e^{-\frac{M_\pi^2(z-1/2)^2}{\frac{8\pi^2}{N_{\rm C}}f_\pi^2}},\nn\\
\frac{4\pi^2f_\pi^2}{N_{\rm C}m^2}&=&\frac{4}{\cN^2 \int dz (z(1-z))^3 \exp\left(-\frac{M_\pi^2(z-1/2)^2}{\frac{4\pi^2}{N_{\rm C}}f_\pi^2}\right)}-1.
\lbqa{piwf5}
Our results are:
\beq
w=0.337\GeV,\;\cN = 13.63,\;m=0.237\GeV.
\lbq{wfpara}
Especially the so obtained value for the mass is very satisfactory because it is very close the constituent quark mass obtained in reference \cite{Dosch:1997II} and used in the photon wavefunction.

By going back to coordinate space we finally obtain the pion wavefunction to be inserted in \eq{profilepi}:
\beqa
\Psi^{\pi^0}_{f h_1 h_2}(\vec{r},z)&=&c_f \int \frac{d^2 p}{(2\pi)^2}e^{i\vec{p}\cdot\vec{r}}\Psi^{\pi^0}_{h_1 h_2}(\vec{p},z)\nn\\
&=&c_f\frac{1}{\sqrt{2}}\left( \de_{+-}-\de_{-+}+2i\frac{r}{m}\left(\de_{++}e^{-i\th}+\de_{--}e^{i\th}\right)\frac{\pa}{\pa r^2}\right)\Psi^{\pi^0}(r,z)\nn\\
{\rm with}\;\Psi^{\pi^0}(r,z)&=&\frac{2\pi f_\pi}{\sqrt{N_{\rm C}}}z(1-z)f(z)e^{-\frac{w^2 r^2}{2}}.
\lbqa{piwf}
Here $\th$ is the angle of $\vec{r}$ in planar polar coordinates. It follows directly from the helicity structure that the $\pi^0$ has no overlap with longitudinal polarized photons. For the transversal case we obtain using \eq{piwf} and the photon wavefunction from reference \cite{Dosch:1997}
\beqa
\sum_{f,h_1, h_2}\Psi^{*\, \pi^0}_{f h_1 h_2}(\vec{r}_\ga,z)\Psi^{\ga}_{f h_1 h_2}(\vec{r}_\ga,z)&=&ie\hat{e}_{\pi^0}f_\pi e^{-\frac{w^2 r_\ga^2}{2}}e^{i\th}z(1-z)f(z)\nn\\
&&\times \left(\ep{\rm K}_1(\ep r_\ga)+\frac{m(Q^2)}{m}r_\ga w^2{\rm K}_0(\ep r_\ga)\right)\nn\\
{\rm with}\;\ep &=&\sqrt{z(1-z)Q^2+m^2(Q^2)}.
\lbqa{pigawf}
Here $e$ is the proton charge, $\hat{e}_{\pi^0}=c_f e_f/e=1/\sqrt{2}$ is the mean charge of the quarks in the pion in units of $e$ and $m(Q^2)$ is the quark mass depending on the virtuality introduced in reference \cite{Dosch:1997II}.

Putting everything together we calculate the differential and total elastic cross section of $\ga p \rightarrow \pi^0 p$. The result for photoproduction ($Q^2=0$) is shown and discussed in \fig{elasticpiresult}.
\betwofi{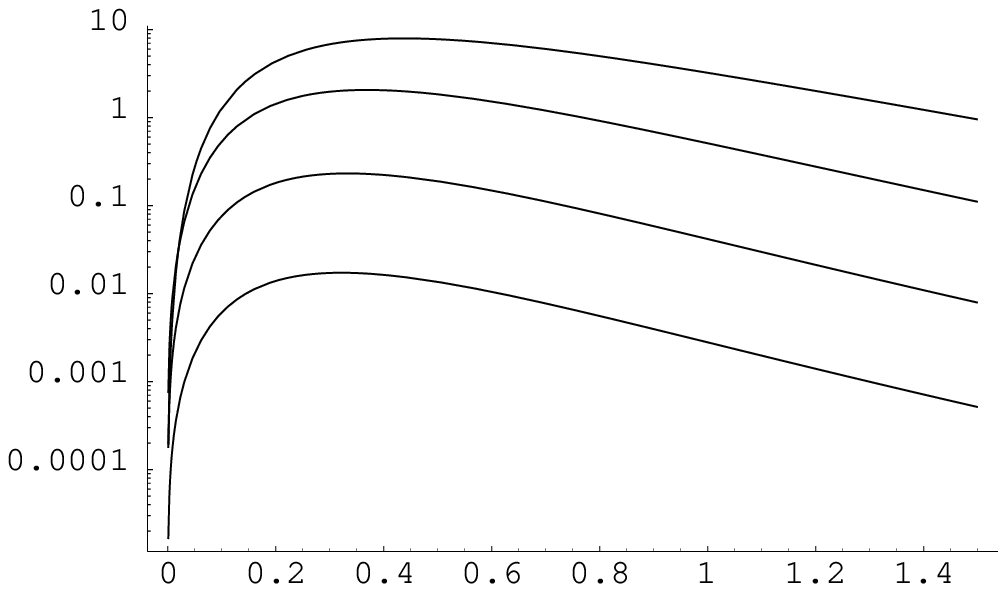}{\vspace*{.8cm}The differential cross section for different diquark sizes}{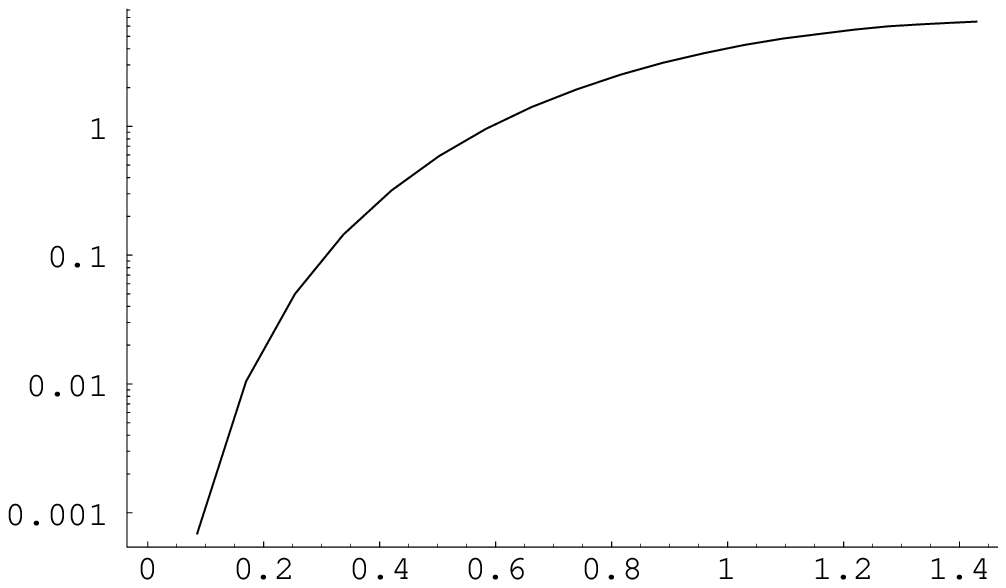}{\vspace*{.8cm}The total elastic cross section as a function of the diquark size}{7.5cm}
\unitlength1cm
\begin{picture}(0,0)
\small
\put(-10.6,5.2){$\frac{\displaystyle{d\si}}{\displaystyle{dt}}[{\rm nb}\GeV^{-2}]$}
\put(-6.3,5.1){$\si [{\rm nb}]$}
\put(-9.9,.6){$t[\GeV^2]$}
\put(-1.0,.6){$d[\fm]$}
\scriptsize
\put(-9.6,4.8){$d=1.472$ fm}
\put(-9.6,4.2){$d=0.662$ fm}
\put(-9.6,3.5){$d=0.338$ fm}
\put(-9.6,2.7){$d=0.170$ fm}
\end{picture}
\lbcap{14cm}{The differential and total elastic cross section for $\ga p \rightarrow \pi^0 p$ with $Q^2=0$. We give the results for different diquark sizes in the proton. The differential cross section is given for diquark sizes $d=0.170$ fm, 0.338 fm, 0.662 fm and 1.472 fm and rises with $d$. For small $d$ the $d^4$ behavior can be seen and the largest size corresponds to a symmetric Mercedes-star like geometry of the proton. By integrating over $t$ we obtain the total elastic cross section which again shows for small $d$ the $d^4$ behavior and saturates for large diquark sizes. Notice that from \mbox{$d=0.5$ fm} to \mbox{$d=0.1$ fm} the cross section drops by three orders of magnitude. For $t=0$ the differential cross section has to vanish due to total angular momentum conservation.}{elasticpiresult}

Increasing $Q^2$ from zero does not change the $t$ dependence of the differential cross section. Only the absolute size of the cross sections drops with increasing virtuality. In \fig{sigelallQ} we show the $d$ dependence of the total elastic cross section for different values of $Q^2$.
\befi{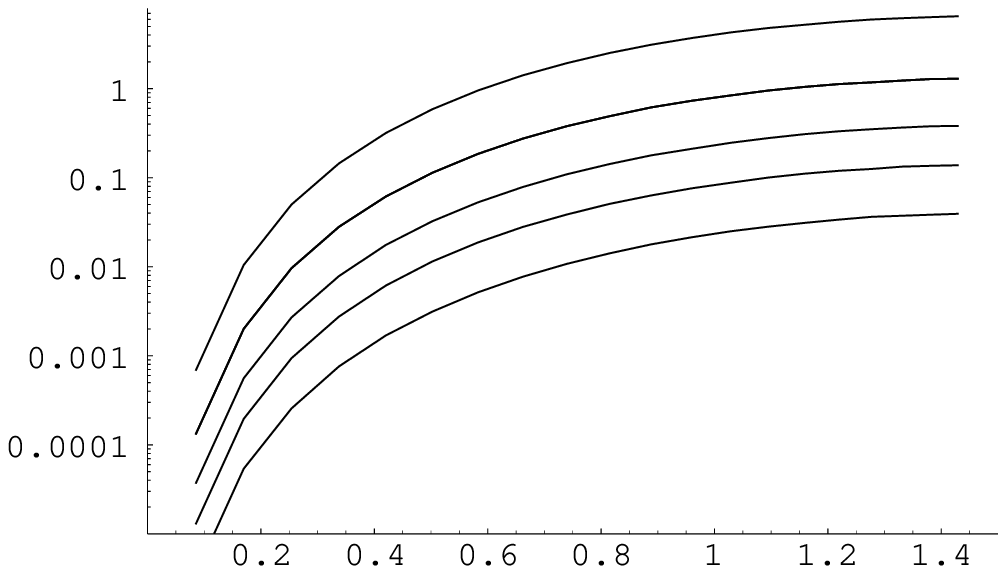}{7.5cm}
\unitlength1cm
\begin{picture}(0,0)
\small
\put(5.5,4.8){$\si [{\rm nb}]$}
\put(10.5,.6){$d[\fm]$}
\scriptsize
\put(11.5,5.0){$Q^2=0$ GeV$^2$}
\put(11.5,4.6){$Q^2=0.5$ GeV$^2$}
\put(11.5,4.2){$Q^2=1.0$ GeV$^2$}
\put(11.5,3.9){$Q^2=2.0$ GeV$^2$}
\put(11.5,3.5){$Q^2=4.0$ GeV$^2$}
\end{picture}
\lbcap{14cm}{The total elastic cross section as a function of the diquark size $d$ for 5 different values of $Q^2$ (0, 0.5, 1.0, 2.0 and 4.0 ${\rm GeV}^2$). With increasing virtuality the cross section decreases.}{sigelallQ}
\subsection{$\pi^0$ production in single dissociation}
In this subsection we calculate the cross section of $\pi^0$ production in single dissociation. A visualization of the process and some definitions are given in \fig{pionsd}.
\befi{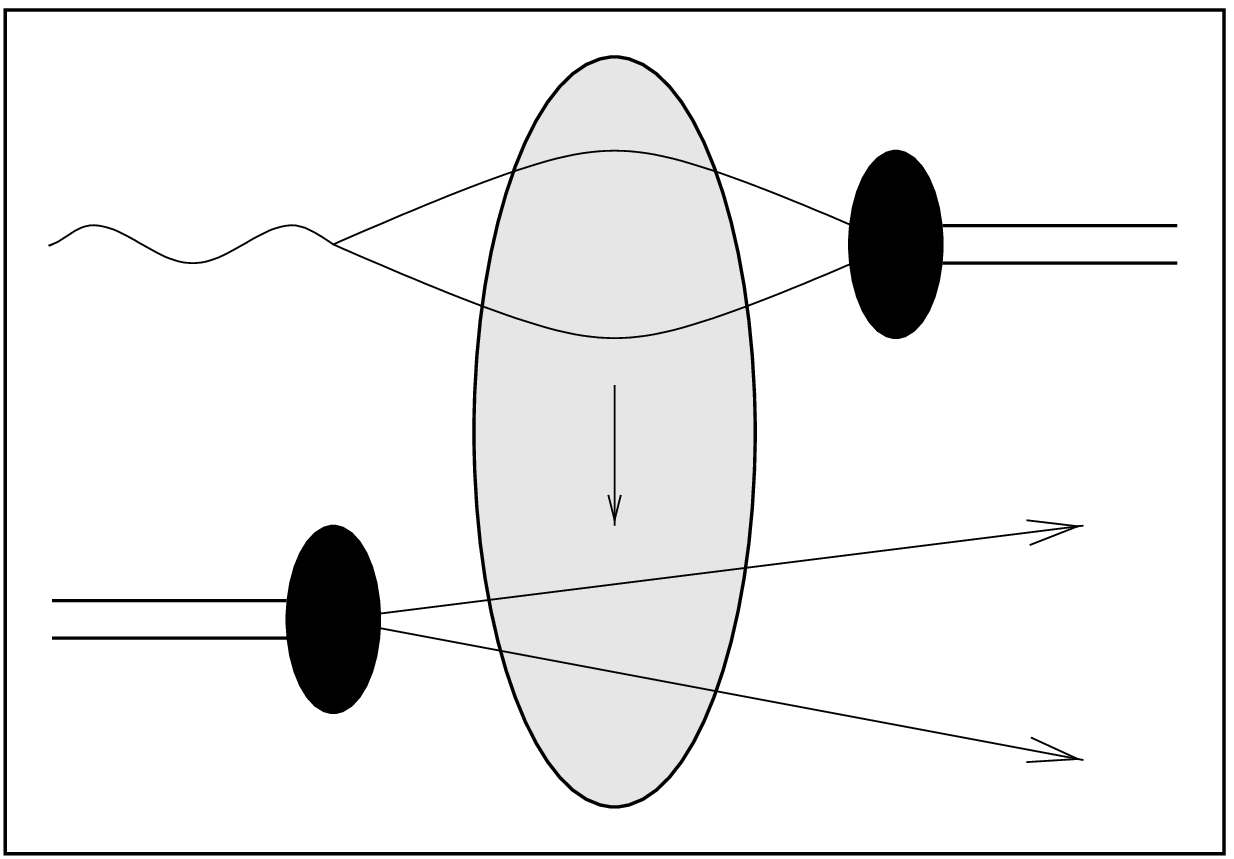}{6cm}
\unitlength1cm
\begin{picture}(0,0)
\small
\put(5.8,4.0){$q$}
\put(5.8,1.6){$p$}
\put(6.9,1.5){$p_2'$}
\put(6.9,2.2){$p_1'$}
\put(10.3,1.2){$p_2$}
\put(10.3,2.4){$p_1$}
\put(8.1,2.6){$\vec{\De}_\perp$}
\put(10.3,4.1){$p_\pi$}
\end{picture}
\lbcap{14cm}{$\pi^0$ production in single dissociation. The black discs denote the phenomenological wavefunctions and the grey disc the nonperturbative diffractive interaction with momentum transfer $\vec{\De}_\perp$. The momenta for the following calculation are defined.}{pionsd}

For single dissociation we have no suppression due to diquark clustering since the overlap of the incoming proton and the outgoing diffractively excited proton will have antisymmetric contributions. We can therefore use the convenient approximation of a proton as a quark-diquark state as discussed for proton-proton scattering in reference \cite{Dosch:1994}. By expanding this reaction in the diquark size $d$ we would find that the leading contribution for single dissociation is of order $\cO (d^0)$. That means the cross sections for single dissociation are rather independent of $d$ as compared to the strong $d^4$ dependence for the elastic case.

In the following we will work with light cone variables. We use the normalization
\[
k=(k^+,k^-,\vec{k}),\;k^2=k^+ k^--\vec{k}^2
\]
and the calculation is done in the leading high-energy limit. The momenta are defined as follows, where we used for the quark and diquark in the final state that $p_1^-+p_2^-=p^-$ in the high-energy limit:
\beqa
q&=&(q^+,-\frac{Q^2}{q^+},\vec{0}),\nn\\
p&=&(\frac{M_{\rm p}^2}{p^-},p^-,\vec{0}),\nn\\
p_1&=&(\frac{\vec{p}_1^2+m^2}{xp^-},xp^-,\vec{p}_1),\nn\\
p_2&=&(\frac{\vec{p}_2^2+m^2}{(1-x)p^-},(1-x)p^-,\vec{p}_2).
\enqa
Then $\vec{p}_1+\vec{p}_2=\vec{\De}_\perp$ and we define the {\it relative momentum} with $\vec{p}_\perp=(1-x)\vec{p}_1-x\vec{p}_2$, which implies $\vec{p}_1-\vec{p}_2=2(\vec{p}_\perp+(x-(1-x))\vec{\De}_\perp)$. In the light cone wavefunction approach this definition of $\vec{p}_\perp$ enters together with $x$ in the wavefunction.

It is easy to see that the light cone wavefunction of the free quark-diquark pair with momenta $p_1$ and $p_2$ in our normalization (see for example \eq{profilepi}) is just a plain wave with an extra factor $\sqrt{x(1-x)}$:
\[
\Ps_{q\bar{q}}(\vec{r},x)=4\pi \sqrt{x(1-x)}e^{i\vec{p}_\perp\cdot \vec{r}}.
\]
For the scattering amplitude we obtain
\beqa
T &=&2is \int  d^2b\,e^{-i\vec{\De}_\perp\cdot\vec{b}}\int \frac{d^2 r_{\rm P}}{4\pi}\Psi^{\rm P}(r_{\rm P})4 \pi \sqrt{x(1-x)}e^{-i\vec{p}_\perp\cdot \vec{r}_{\rm P}}\nn\\
&&\times\int \frac{d^2 r_\ga}{4\pi}dz\sum_{f,h_1, h_2}\Psi^{*\, \pi^0}_{f h_1 h_2}(\vec{r}_\ga,z)\Psi^{\ga}_{f h_1 h_2}(\vec{r}_\ga,z) \tilde{J}^{C=-1}_{\rm DD}.
\lbqa{Tpionsd}
The differential cross section is given by
\beq
d\si = \frac{(2\pi)^4}{2s}|T|^2 dR_3,
\lbq{sipionsd}
where $dR_3$ is the five dimensional phase space
\[
dR_3=\de^4(q+p-p_1-p_2-p_\pi)\de(p_1^2-m^2)\de(p_2^2-m^2)\de(p_\pi^2-m_\pi^2)\frac{d^4p_1 d^4p_2 d^4p_\pi}{(2\pi)^9}
\]
which reduces in the high-energy limit to
\beq
dR_3=\frac{1}{4s}\frac{1}{(2\pi)^9}\frac{1}{x(1-x)}dx d^2p_\perp d^2\De_\perp .
\lbq{dr3}
We have
\[
M^2=(p_1+p_2)^2=\frac{\vec{p}^2_\perp+m^2}{x(1-x)}.
\]
Putting \eq{Tpionsd}, \eq{sipionsd} and \eq{dr3} together and integrating over $\De_\perp$ we obtain for the cross section differential in $M^2$
\beqa
\frac{d\si}{d M^2}&=&\int dx \,d\th_{p_\perp} \int  \frac{d^2 b}{2\pi} x(1-x) \Bigg| \int \frac{d^2 r_{\rm P}}{4\pi}\Psi^{\rm P}(r_{\rm P})e^{-i\sqrt{M^2x(1-x)}r_{\rm P}\cos (\th_{p_\perp}-\th_{r_{\rm P}})} \nn\\
&&\times \int \frac{d^2 r_\ga}{4\pi}dz\sum_{f,h_1, h_2}\Psi^{*\, \pi^0}_{f h_1 h_2}(\vec{r}_\ga,z)\Psi^{\ga}_{f h_1 h_2}(\vec{r}_\ga,z) \tilde{J}^{C=-1}_{\rm DD} \Bigg|^2.
\lbqa{dsidMpionsd}
Here $\th_{p_\perp}$ and $\th_{r_{\rm P}}$ denote the angle of $\vec{p}_\perp$ and $\vec{r}_{\rm P}$ in polar coordinates. As in the elastic case only transversal polarized photons contribute (\eq{pigawf}).\\
In our nonperturbative model only values of $M^2$ which are not too large as compared to the nonperturbative scale contribute. Integrating $M^2$ up to infinity we obtain 
\beq
\si=\int \frac{d^2 r_{\rm P}}{4\pi}dx\left| \Psi^{\rm P}(r_{\rm P})\right|^2\int  d^2b\left| \int \frac{d^2 r_\ga}{4\pi}dz\sum_{f,h_1, h_2}\Psi^{*\, \pi^0}_{f h_1 h_2}(\vec{r}_\ga,z)\Psi^{\ga}_{f h_1 h_2}(\vec{r}_\ga,z) \tilde{J}^{C=-1}_{\rm DD} \right|^2.
\lbq{sitotpionsd}
Putting as an approximation $x=1/2$ instead of integrating over the full range of $x$ has only a small numerical effect because $\tilde{J}^{C=-1}_{\rm DD}$ depends only weakly on $x$.

In \fig{logdsigdm} we give as a function of $Q^2$ the total cross section (\eq{sitotpionsd}) as compared to the elastic case with a diquark size $d=1.472\fm$ (Mercedes-star geometry, see \fig{sigelallQ}) and $d=0.338\fm$ (maximal diquark size which is compatible with the odderon contribution to $p$-$\bar p$ scattering). In \fig{verh} we show that integrating the differential cross section over $M^2$ up to ${\hat M}^2 = 2\GeV^2$ and ${\hat M}^2 = 3\GeV^2$ gives for all values of $Q^2$ already about 90\% and 95\% of the total cross section respectively.
\befi{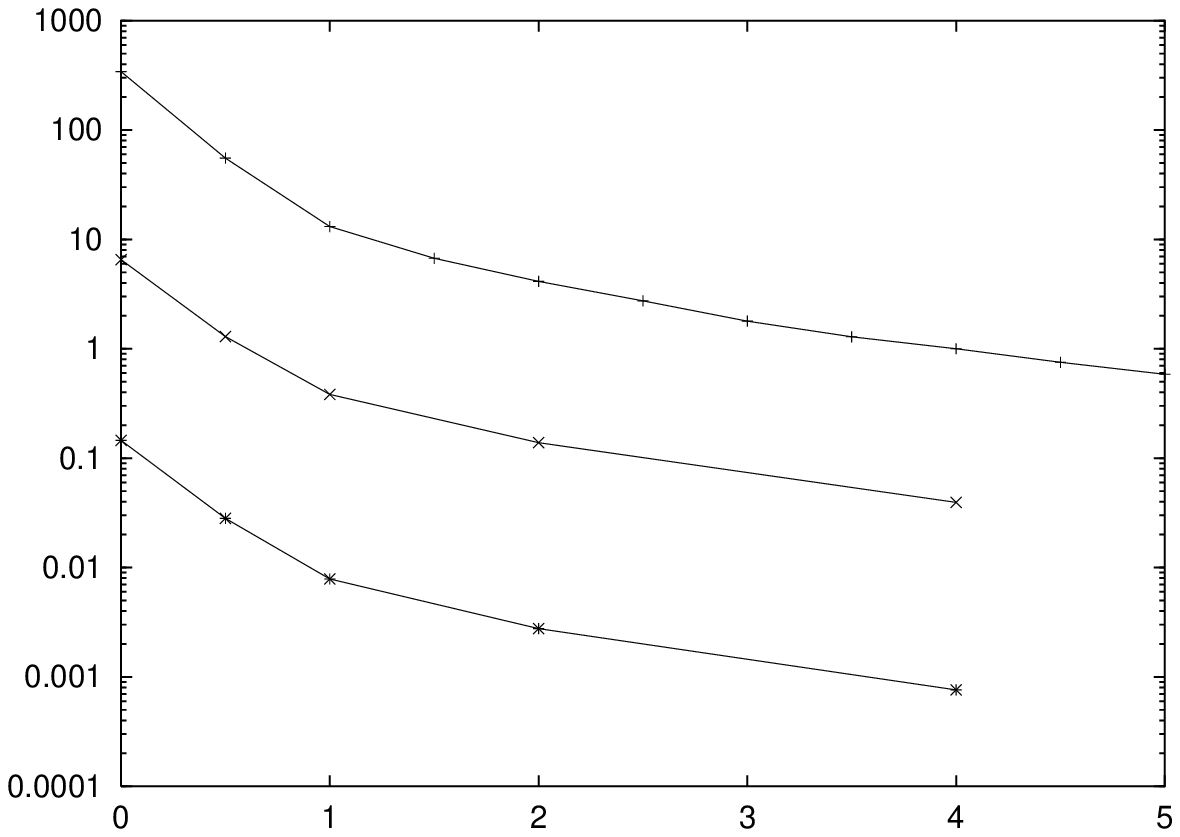}{10cm}
\unitlength1cm
\begin{picture}(0,0)
\small
\put(+11.,.8){$Q^2[\GeV^2]$}
\put(+5.0,7.0){$\si [{\rm nb}]$}
\scriptsize
\put(+10.,5.3){proton dissociation}
\put(+8.5,4.3){elastic scattering, $d=1.472$ fm}
\put(+8.5,2.8){elastic scattering, $d=0.338$ fm}
\end{picture}
\lbcap{14cm}{The total cross section (\eq{sitotpionsd}) as a function of $Q^2$ (upper curve) as compared to the elastic case with $d=1.472$ fm (Mercedes-star geometry, see \fig{sigelallQ}) and $d=0.338$ fm (maximal diquark size which is compatible with the odderon contribution to $p$-$\bar p$ scattering). The elastic cross section is at least a factor of 50 smaller. The curve also shows that the asymptotic $Q^2$ dependence is not reached. For $Q^2\approx 2\,{\rm GeV}^2$ the cross section goes like $1/Q^4$ and for larger virtualities like $1/Q^5$.}{logdsigdm}
\befi{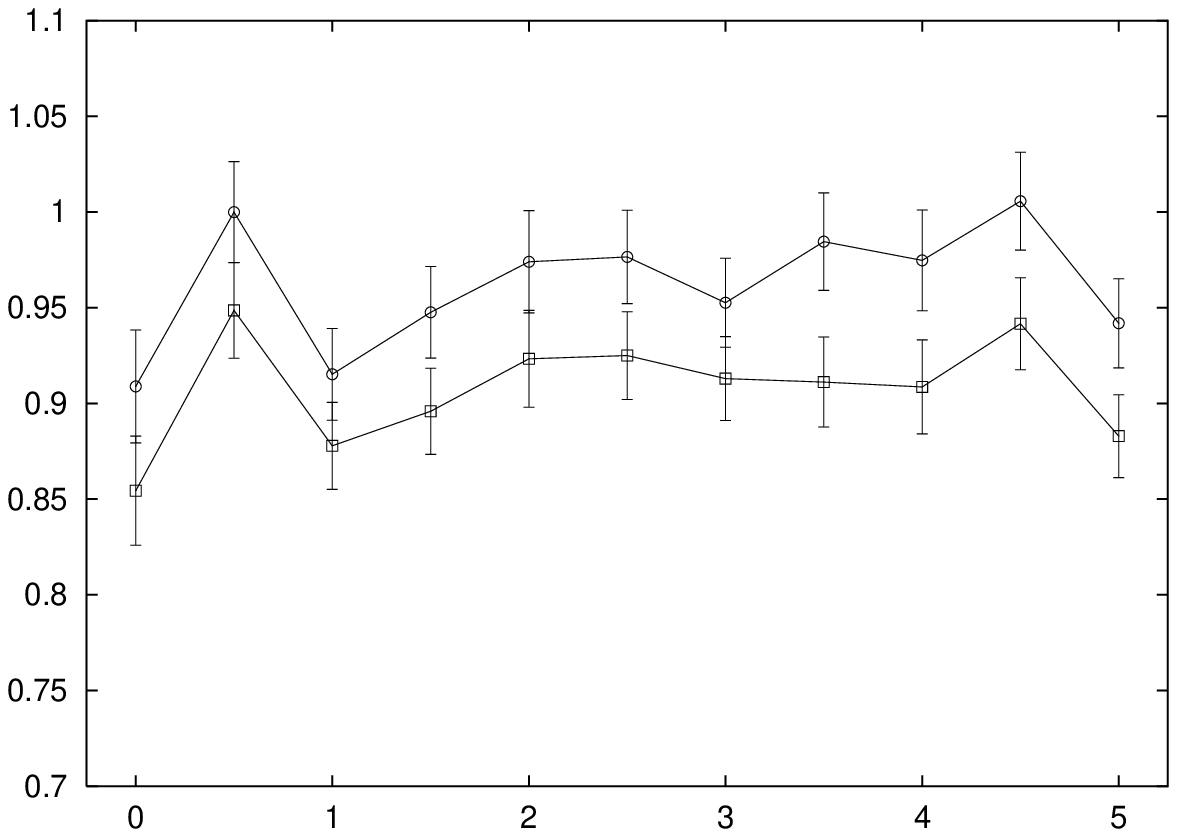}{10cm}
\unitlength1cm
\begin{picture}(0,0)
\small
\put(+11.,.7){$Q^2[\GeV^2]$}
\put(+4.5,6.8){$\int_0^{\hat M^2}\frac{\displaystyle{d\si}}{\displaystyle{d M^2}} d M^2 / \si$}
\end{picture}
\lbcap{14cm}{The differential cross section (\eq{dsidMpionsd}) integrated for \mbox{$M^2\le {\hat M}^2$ with ${\hat M}^2 = 2 \,{\rm GeV}^2$} and \mbox{${\hat M}^2 = 3\,{\rm GeV}^2$} divided by the total cross section (\eq{sitotpionsd}). This shows that $\hat M^2 =2\,{\rm GeV}^2$ already gives about 90 percent and $\hat M^2 =3\,{\rm GeV}^2$ about 95 percent of the total cross section. The errors indicate our numerical accuracy which is about 3 percent for the cross sections.}{verh}

Our results show, that the $\pi^0$ production in single dissociation is much bigger (by a factor about 50) then the elastic case even if there is no suppression due to diquark clustering in the proton.
\section{Summary}
If the odderon contribution to scattering amplitudes is suppressed as is strongly indicated by the good agreement between $p$-$p$ and $\bar p$-$p$ scattering and if this suppression is due to diquark clustering we predict a strong suppression of all processes dominated by odd $C,P$ exchange if the target proton is not dissociated. 
We have shown that a promising  process to isolate the odderon contribution is photoproduction of pions with single dissociation, i.e.~breakup of the target proton. In our model the cross section for this process is about 300 nbarn and thus about fifty times larger than elastic photoproduction, even if no suppression due to diquark clustering takes place. Final neutrons in the dissociation channel could be an easy signature for the breakup. For realistic comparisons with experimental data off course the kinematic cuts and the interference with photon exchange has to be taken into account (see reference \cite{Kilian:1997}).  
\section*{Acknowledgments}
We want to thank Wolfgang Kilian and Karlheinz Meier for fruitful discussions. We also thank D.Y.~Ivanov for pointing out an error of our calculation in an intermediate stage. One of us (M.R.) is grateful to the {\it MINERVA-Stiftung} for financial support.
\clearpage
\section*{Appendix: Expansion of the $C=-1$ profile function in the diquark size of the baryon}
In this appendix we calculate $\tilde{J}_{\rm DB}^{C=-1}$ by expanding the Wegner-Wilson loop construction of the baryon (\eq{Jtilde}) in the dipole size. We put the arbitrarily placed common line of the three Wegner-Wilson loops on the position of quark 2 (see \fig{3loops}). Therefore \mbox{$\bW_{b'b}[S_{22}]\Rightarrow \de_{b'b}$} and $B_2$ becomes
\beqa
B_2 &=& \frac{1}{6}\ep_{abc}\ep_{a'b'c'}\left\{ \bW_{a'a}[S_{21}]\de_{b'b}\bW_{c'c}[S_{23}]-\de_{a'a}\de_{b'b}\de_{c'c}\right\} \nn\\
&=&\frac{1}{6}\left\{ \bW_{aa}[S_{21}]\bW_{cc}[S_{23}]-\bW_{ca}[S_{21}]\bW_{ac}[S_{23}]-6\right\} \nn\\
&=&\frac{1}{6}\left\{ \Tr\bW[S_{21}]\Tr\bW[S_{23}]-\Tr\left\{\bW[S_{21}]\bW[S_{23}]\right\}-6\right\},
\lbqa{WWloopa2}
where $S_{21}$ is a loop connecting quark 1 with quark 2 of the diquark, whereas $S_{23}$ a small loop of extension $d$ connecting the two quarks of the diquark. We use this $B_2$ to calculate the profile function (\eq{redampl}). For the dipole we have \eq{WWloopb} and we have to expand $W_1$ and $B_2$ up to third order in the field strengths to obtain the $C=-1$ contribution. Expanding the large loop up to the third order we obtain the result for a point like diquark, which is canceled by the symmetric baryon wavefunction. Expanding the small loop up to third order is proportional to $d^3$ and can be neglected. Expanding one loop to second and the other to first order contributes only to the second term in the sum of \eq{WWloopa2}. Expanding the small loop only up to first order gives a contribution which is of $\cO (d^1)$. But we will show that the $\cO(d^1)$ contribution is canceled because of symmetry arguments. For $\tilde{J}_{\rm DB}^{C=-1}$ we obtain in leading order in $d$
\[
\tilde{J}_{\rm DB}^{C=-1}=\frac{1}{3\times 6}<\Tr \bW [S_1]\Tr\left\{\bW[S_{21}]\bW[S_{23}]\right\}> + \cO (d^3),
\]
where the third order in the field strengths of both traces has to be calculated. The result is:
\beqa
\tilde{J}_{\rm DB}^{C=-1}&=& -i\frac{5}{8N_{\rm C}^2(N_{\rm C}^2-1)^2 12^3 36}\frac{3}{2}\left((\tilde{\ch}_{12}-\tilde{\ch}_{13}-\tilde{\ch}_{22}+\tilde{\ch}_{23})(\tilde{\ch}_{11}-\tilde{\ch}_{12}-\tilde{\ch}_{21}+\tilde{\ch}_{22})^2\right. \nn\\
&&-(\tilde{\ch}_{12}-\tilde{\ch}_{13}-\tilde{\ch}_{22}+\tilde{\ch}_{23})^2(\tilde{\ch}_{11}-\tilde{\ch}_{12}-\tilde{\ch}_{21}+\tilde{\ch}_{22})\Big)+ \cO (d^3).
\lbqa{od2e}
The first term of the sum results from the second order expansion of the large loop. We will show that the $\cO(d^1)$ contribution is canceled by the wavefunction integration and we have to calculate the $\cO (d^2)$ contribution and add it to the second term in \eq{od2e} which comes from the second order expansion of the small loop and is of $\cO (d^2)$ anyhow.

To finish the expansion we introduce the symmetric ``point'' 4, which is in the middle of the diquark, just opposite of quark 1 of the baryon (see \fig{symmetry}). Then we have with
\[
a=(\tilde{\ch}_{12}-\tilde{\ch}_{13}-\tilde{\ch}_{22}+\tilde{\ch}_{23}),\,b=(\tilde{\ch}_{11}-\tilde{\ch}_{12}-\tilde{\ch}_{21}+\tilde{\ch}_{22})
\]
the following expansion for the first term in \eq{od2e}:
\beqa
ab^2&=&a((\tilde{\ch}_{11}-\tilde{\ch}_{14}-\tilde{\ch}_{21}+\tilde{\ch}_{24})+(\tilde{\ch}_{14}-\tilde{\ch}_{24}-\tilde{\ch}_{12}+\tilde{\ch}_{22}))^2\nn\\
&=&a(\tilde{\ch}_{11}-\tilde{\ch}_{14}-\tilde{\ch}_{21}+\tilde{\ch}_{24})^2+a(\tilde{\ch}_{14}-\tilde{\ch}_{24}-\tilde{\ch}_{12}+\tilde{\ch}_{22})^2+\nn\\
&&2a(\tilde{\ch}_{11}-\tilde{\ch}_{14}-\tilde{\ch}_{21}+\tilde{\ch}_{24})(\tilde{\ch}_{14}-\tilde{\ch}_{24}-\tilde{\ch}_{12}+\tilde{\ch}_{22}).
\lbqa{ex1}
The first term in the sum is of order $\cO(d^1)$ but this contribution vanishes by integrating over the wavefunctions. This can be seen by rotating the two objects such, that $\theta$ is replaced by $2\pi-\th$ (see \fig{symmetry}).
\betwofi{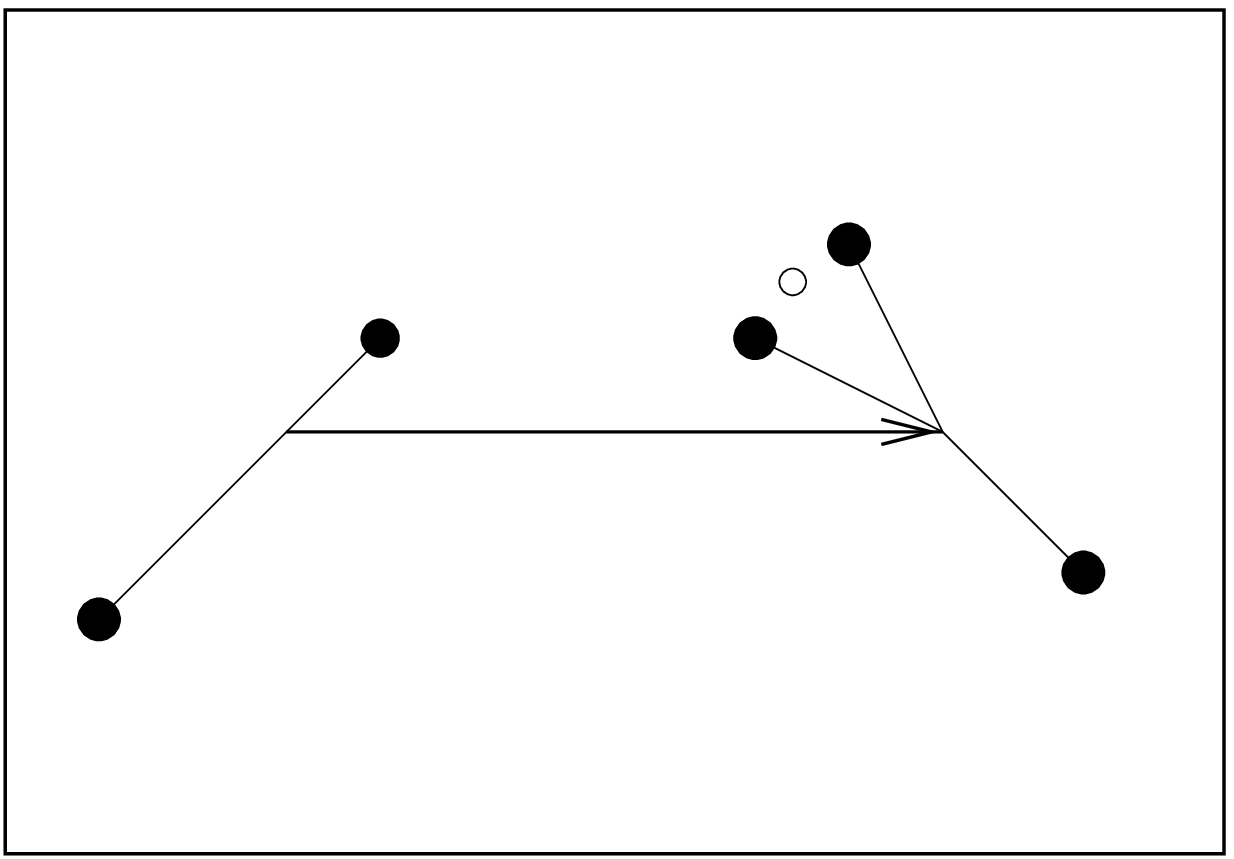}{}{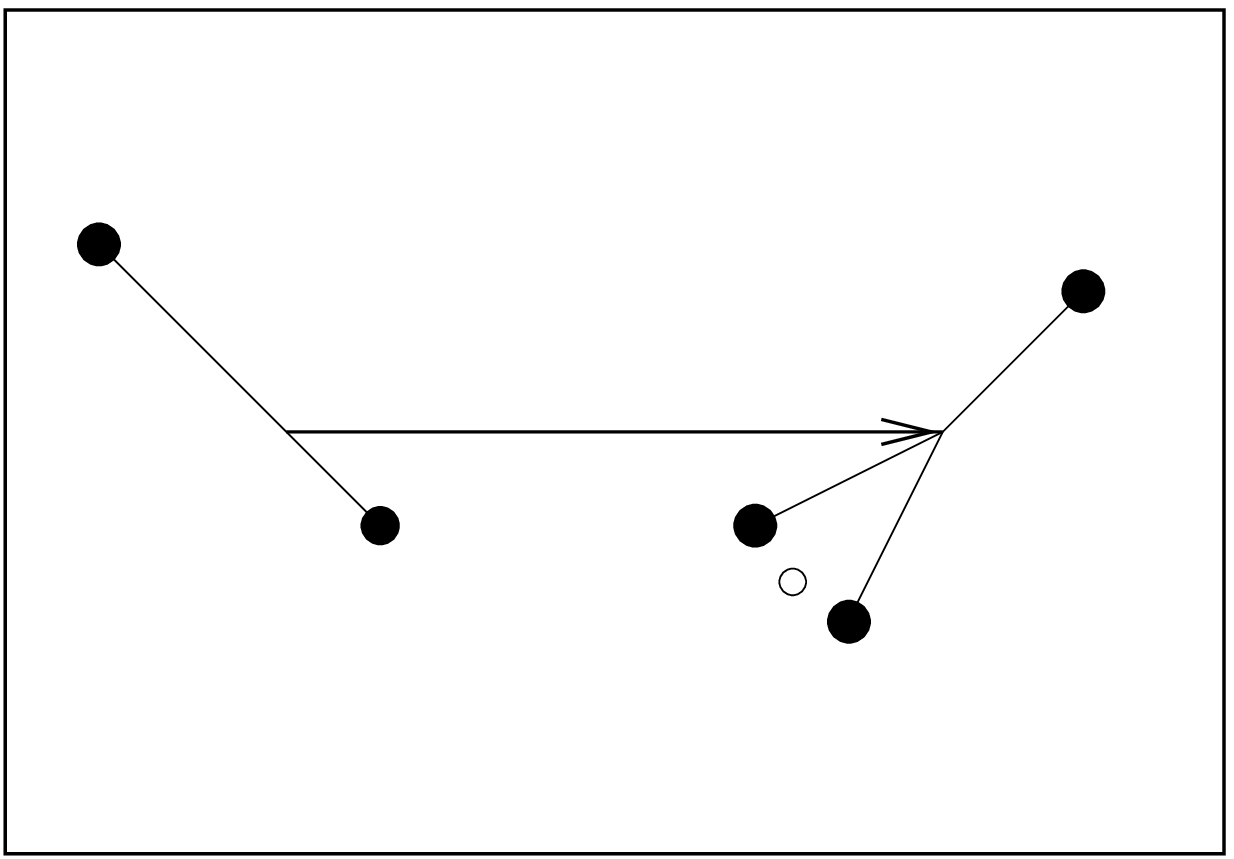}{}{6cm}
\unitlength1cm
\begin{picture}(0,0)
\small
\put(-13.1,1.8){$\vec{b}$}
\put(-3.2,1.8){$\vec{b}$}
\put(-.5,3.0){1}
\put(-2.7,1.5){2}
\put(-2.1,.9){3}
\put(-4,1.5){1}
\put(-5.8,3.4){2}
\put(-10.5,1.2){1}
\put(-12.6,2.9){3}
\put(-12.1,3.4){2}
\put(-14,2.9){1}
\put(-15.8,1.1){2}
\end{picture}
\lbcap{14cm}{Illustration of the rotation $\th\Rightarrow 2\pi-\theta$. The dots denote the (anti)quark positions and the open dot the symmetric ``point'' 4 introduced in the text. The numbers belong to the index of $\tilde{\ch}_{ij}$.}{symmetry}

By this rotation the baryon wavefunction and $(\tilde{\ch}_{11}-\tilde{\ch}_{14}-\tilde{\ch}_{21}+\tilde{\ch}_{24})$ is unchanged but $(\tilde{\ch}_{12}-\tilde{\ch}_{13}-\tilde{\ch}_{22}+\tilde{\ch}_{23})$ changes its sign because $\tilde{\ch}_{_{i2}}\Leftrightarrow \tilde{\ch}_{_{i3}}$. So, besides the dipole wavefunction we have a change of the sign. The antisymmetric dipole wavefunction has a factor $e^{i\theta}=\cos\th+i\sin\th$. The $\cos\th$ is symmetric under $\th\Rightarrow 2\pi-\theta$ from which follows that the first term in the sum of \eq{ex1} is canceled for this part of the dipole wavefunction. To finish the proof of the cancellation of the $\cO(d^1)$ we show finally that the $\sin\th$ part of the dipole wavefunction does not contribute to the amplitude anyhow. Therefore we consider in the profile function (\eq{dbcminus}) a rotation of the dipole where $\th\Rightarrow \pi-\theta$, replace $z\Rightarrow (1-z)$ and integrate over the baryon orientation (see \fig{symmetry2}).
\betwofi{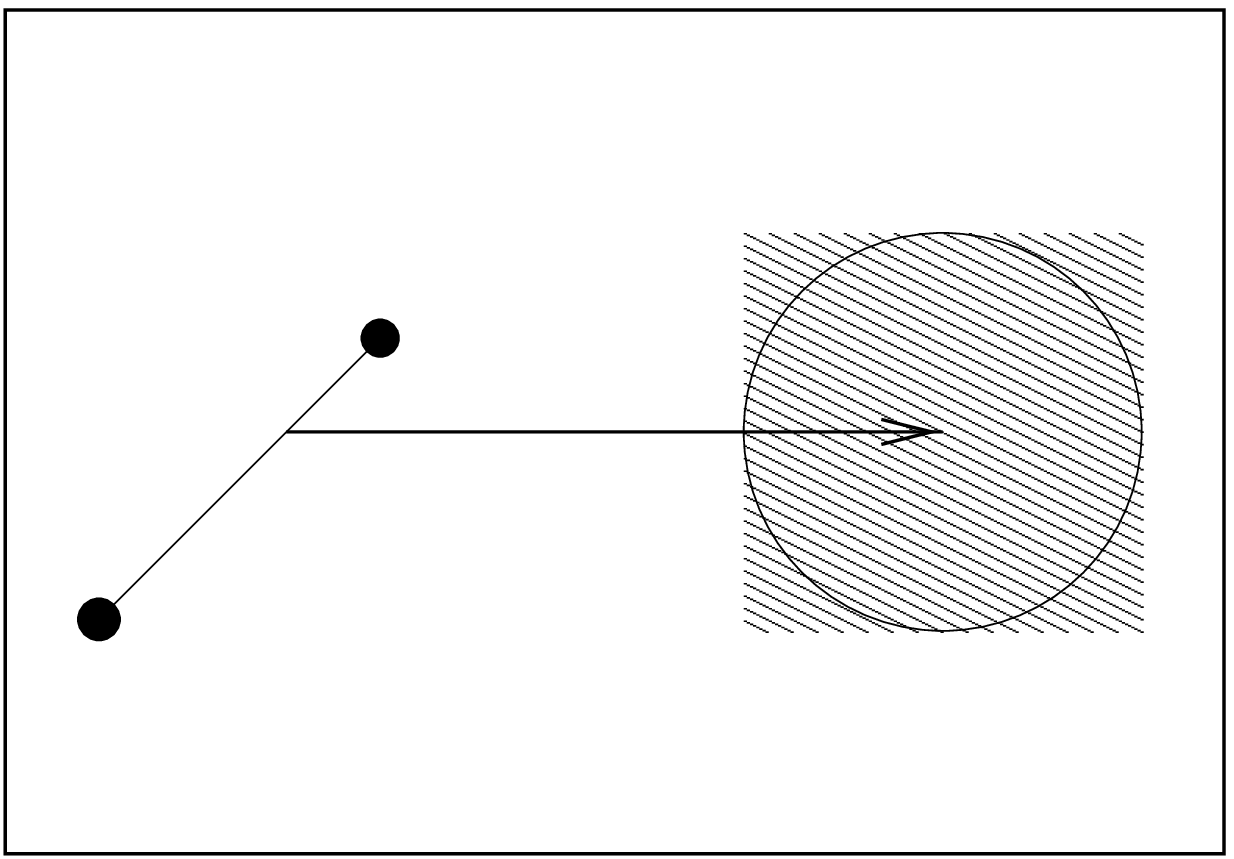}{}{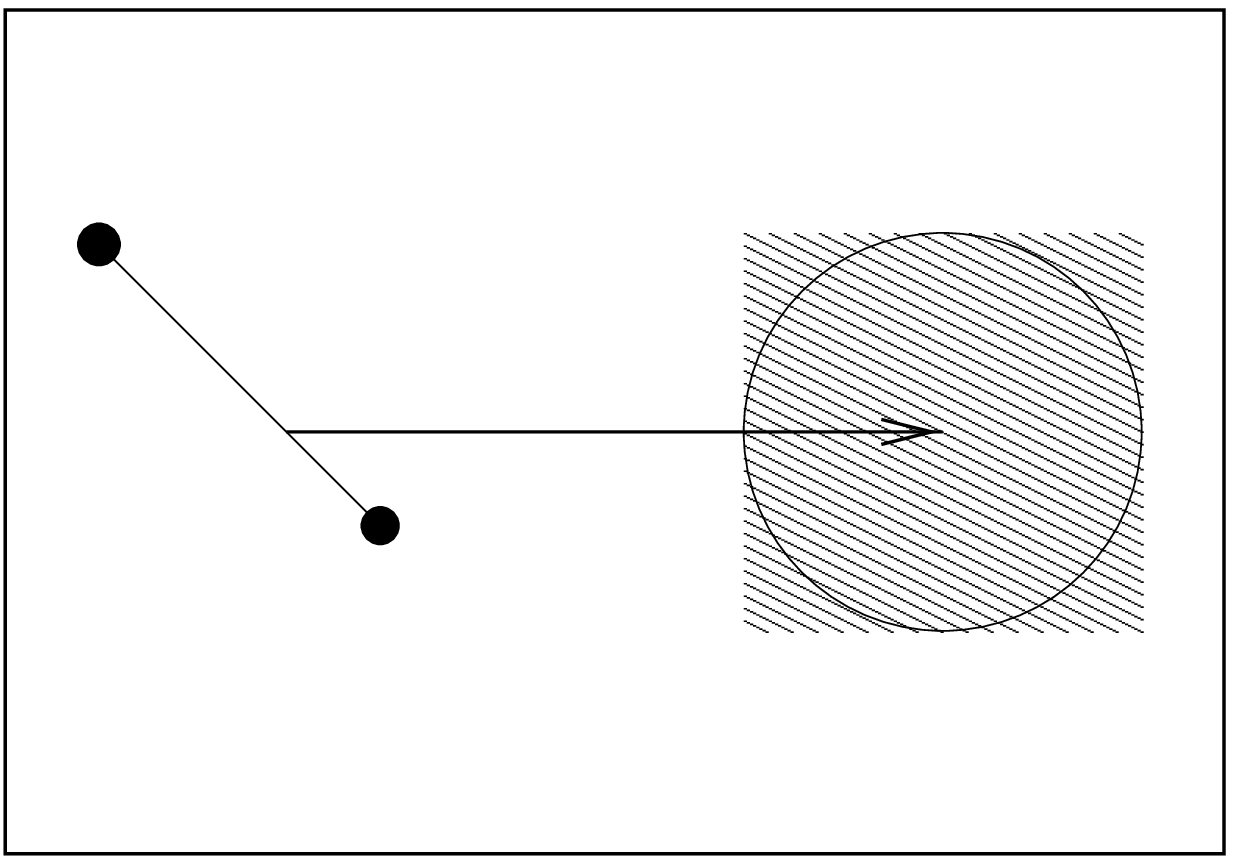}{}{6cm}
\unitlength1cm
\begin{picture}(0,0)
\small
\put(-13.1,1.8){$\vec{b}$}
\put(-3.2,1.8){$\vec{b}$}
\put(-4,1.5){2}
\put(-5.8,3.4){1}
\put(-14,2.9){1}
\put(-15.8,1.1){2}
\end{picture}
\lbcap{14cm}{Illustration of the rotation $\th\Rightarrow \pi-\theta$ and $z\Rightarrow (1-z)$. The dots denote the (anti)quark positions and the grey disk the baryon integrated over its orientation. The numbers belong to the index of $\tilde{\ch}_{ij}$.}{symmetry2}

The $\sin\th$ part of the dipole wavefunction is symmetric under this manipulations but the profile function integrated over the baryon orientation changes its sign. So, the integration over the dipole wavefunction cancels this contribution to the amplitude.

The second term in \eq{ex1} is of $\cO (d^3)$ and can be neglected. Using
\[
(\tilde{\ch}_{11}-\tilde{\ch}_{14}-\tilde{\ch}_{21}+\tilde{\ch}_{24})=b+\cO (d^1),\,(\tilde{\ch}_{14}-\tilde{\ch}_{12}-\tilde{\ch}_{24}+\tilde{\ch}_{22})=-\frac{1}{2}a+\cO (d^2)
\]
the third term in \eq{ex1} becomes $-a^2b+\cO (d^3)$ and we obtain finally for \eq{od2e}:
\beqa
\tilde{J}_{\rm DB}^{C=-1}&=& -i\frac{5}{8N_{\rm C}^2(N_{\rm C}^2-1)^2 12^3 36}\frac{3}{2}\left(-a^2 b-a^2 b\right)+\cO (d^3)\nn\\
&=&-i\frac{5}{8N_{\rm C}^2(N_{\rm C}^2-1)^2 12^3 36}(-3)a^2b+\cO (d^3)\nn\\
&=&-i\frac{5}{8 N_{\rm C}^2(N_{\rm C}^2-1)^2 12^3 36}\nn\\
&&\times\left( (-3)\left(\tilde{\ch}_{11}-\tilde{\ch}_{12}-\tilde{\ch}_{21}+\tilde{\ch}_{22}\right)\left(\tilde{\ch}_{12}-\tilde{\ch}_{13}-\tilde{\ch}_{22}+\tilde{\ch}_{23}\right)^2\right)\nn\\
&&+\cO(d^3).
\enqa
This is the leading contribution to the profile function proportional to $d^2$ resulting in a differential cross section proportional to $d^4$.

\end{document}